\documentstyle[twocolumn,aps,floats,epsf]{revtex}
\begin{document}
\title{Kondo Effect in Systems with Spin Disorder}
\author{Igor E. Smolyarenko$^a$ and Ned S. Wingreen$^b$}
\address{$^a$Cavendish Laboratory, University of Cambridge,
Cambridge CB3 0HE, United Kingdom \\ $^b$NEC Research
Institute, 4 Independence Way, Princeton, NJ 08540}
\date{\today}
\draft
\maketitle

\begin{abstract}
We consider the role of static disorder in the spin sector
of the one- and two-channel Kondo models. The distribution
functions of the disorder-induced effective energy splitting
between the two levels of the Kondo impurity are derived to
the lowest order in the concentration of static
scatterers. It is demonstrated that the distribution
functions are strongly asymmetric, with the typical
splitting being parametrically smaller than the average rms
value. We employ the derived distribution function of
splittings to study the temperature dependence of the
low-temperature conductance of a sample containing an
ensemble of two-channel Kondo impurities. The results are
used to analyze the consistency of the two-channel Kondo
interpretation of the zero-bias anomalies observed in
Cu/(Si:N)/Cu nanoconstrictions.
\end{abstract}

\section{Introduction}

The Kondo effect, that is the low-temperature screening of
dynamical quantum defects in metals by band electrons, has
been extensively studied during the past thirty years and is
by now well understood (see, {\em e.g.},
Refs.~\onlinecite{Kondo,Hewson}). The continuing interest in
the problem is motivated chiefly by the search for novel
realizations of the effect~\cite{cox}. For example, {\em
local} non-Fermi-liquid ground states have been predicted
theoretically~\cite{NB} for certain types of dynamical
defects coupled to band electrons. The behavior of these
impurities is sometimes invoked in efforts to understand the
non-Fermi-liquid behavior of strongly correlated systems,
such as heavy fermion materials and high-temperature
superconductors~\cite{Hewson}.

One of the main difficulties encountered in the
interpretation of experimental data from novel Kondo systems
is the fact that defects with internal degrees of freedom
very seldom represent the only type of disorder in the
system. More often, a considerable number of random static
defects affecting band electrons are also
present. Scattering of electrons on this static disorder may
alter the experimental signature of the Kondo 
effect~\cite{Fischer,ohkawa,Vladar,suga,FB,Kotliar,WAM,Phillips},
sometimes masking genuine non-Fermi-liquid behavior, or,
vice versa, possibly mimicking it in Fermi-liquid 
systems~\cite{moreKotliar}.

The Kondo effect requires that at least two internal states
of the impurity be degenerate or, at least, that their
energy difference be much smaller than the Kondo temperature
$T_K$. Barring the cases of accidental degeneracy between
the states of the dynamical impurity, degeneracy occurs as a
consequence of a symmetry, {\em e. g.}, invariance under
time-reversal transformation in the case of the magnetic
one-channel Kondo effect. Accordingly, the possible types of
disorder can be separated into two classes, depending on
whether disorder destroys the relevant symmetry of the
Hamiltonian.

A typical example of the first class is a dilute magnetic
alloy with a finite concentration of non-magnetic
defects. Assuming that the charge state of the magnetic
impurity does not change as a result of the interaction with
conduction electrons (thus excluding the mixed-valence
regime of the Anderson model), potential scattering on
static disorder does not involve the degree of freedom of
conduction electrons which is coupled to the magnetic
impurity -- their spin. The Hamiltonian remains invariant
under time reversal, and the spin states of the impurity are
degenerate even in the presence of disorder. The Kondo
temperature $T_K$, on the other hand, is affected by
potential scattering of electrons. The nature of the ground
state in systems of this type has been studied in
Refs.~\cite{ohkawa,FB,Kotliar,moreKotliar}.

Another example of such symmetry-preserving disorder is
given by spin-orbit scattering~\cite{MW,UZG} in magnetic
Kondo systems. The corresponding Hamiltonian is also
invariant under time reversal, and Kramers' theorem ensures
that each orbital state of conduction electrons is doubly
degenerate, so their coupling to magnetic impurities does
not lift the degeneracy of impurity states.

An entirely different situation is encountered when
scattering on static defects breaks the relevant
symmetry. Kondo coupling between the band electrons and the
dynamic defect then leads to symmetry-breaking contributions
to the self-energy of the dynamic defect. The
frequency-independent part of these contributions can be
reinterpreted as an extra term in the {\em bare} Hamiltonian
of the dynamic defect, and the difference between its
eigenvalues (the energy difference between the ``up'' and
``down'' states of the defect) is an effective splitting
$\Delta$. Being induced by random scattering, the splitting
itself is a random quantity. This type of model was studied,
for example, in Ref.~\onlinecite{MW}, where the {\em
average} splitting induced by broken time-reversal
invariance due to the combined effect of random spin-orbit
scattering and weak magnetic field was computed. A similar
model has been encountered in the study of internal magnetic
field distributions in spin glasses~\cite{HK}.

The present study of models of this type has been motivated
in part by the discussion in Ref.~\onlinecite{WAM} of
different theoretical interpretations of zero-bias anomalies
in Cu/Si:N/Cu
nanoconstrictions~\cite{RB}. The zero-bias anomalies first
reported in Ref.~\onlinecite{RB} were observed in
nanoconstrictions formed by etching a bowl-shaped cavity in
an insulating $\mathrm{Si}_3\mathrm{N}_4$ substrate before
covering both sides with vacuum-deposited Cu films. The
anomalies are characterized by a $\sqrt{V}$ dip in
conductance $G$ at small bias $V$, a corresponding
$\sqrt{T}$ temperature dependence of conductance at $V=0$,
and, more generally, a scaling function of the form
$G\left(V,T\right)-G\left(0,T\right)=
T^{1/2}\Gamma\left(V/T\right)$ where $\Gamma\left(x\gg
1\right)\propto x^{1/2}$.  These features were interpreted
in Refs.~\onlinecite{RB,RBLvD} (see also
Ref.~\onlinecite{RB2} for additional experimental results
and Ref.~\onlinecite{vonDelft&Co} for a comprehensive
review) as consistent with the scaling properties at low
temperatures of the two-channel Kondo model~\cite{NB}. The
observed absence of Zeeman splitting led to a
conclusion~\cite{RB} that a non-magnetic realization of the
two-channel Kondo model of the type suggested by Vlad\'{a}r and
Zawadowski~\cite{Zaw} might be responsible for the observed
anomaly.

The two-channel Kondo model, first proposed in
Ref.~\onlinecite{NB} (see also Ref.~\onlinecite{CoxZaw} for
an extensive review) to classify magnetic properties of rare
earth materials, is characterized by doubling of the degrees
of freedom of conduction electrons as compared to the usual
one-channel case, while the dynamic impurity is still a
``spin-up, spin-down'' doublet. In other words, each orbital
state of conduction electrons acquires, in addition to its
spin, an extra label, ``flavor'', which is silent in the
sense that the scattering on the dynamic impurity conserves
the flavor quantum number. Even so, the strongly correlated
ground state of this model has been predicted to exhibit
unusual and rather distinctive scaling properties, markedly
different from the Fermi-liquid-like ground-state properties
of ordinary Kondo impurities. 

In the original model proposed in Ref.~\onlinecite{NB}, the
flavor degrees of freedom were constructed out of different
angular momentum states of conduction electrons.
Subsequently, it was proposed in a series of papers by
Vlad\'{a}r and Zawadowski~\cite{Zaw} that an {\em effective}
two-channel Kondo model may emerge in an entirely different
context, where the roles of orbital angular momentum and
spin of conduction electrons are interchanged. The role of a
dynamic impurity in such non-magnetic realizations of the
Kondo effect is assumed to be played by a two-level system
(TLS) -- an atom or a group of atoms tunneling between two
nearly degenerate states. If transitions between the two
states of the TLS involve a transfer of charge, the
transition amplitude becomes dependent on the density of
conduction electrons via the Coulomb interaction~\cite{Zaw}.
The parity of electronic states with respect to the center
of the spatially extended defect becomes the active degree
of freedom -- ``pseudospin''. The physical spin assumes the
role of the silent ``flavor'' degree of freedom, providing
two independent channels (in the absence of spin scattering)
for the screening of the dynamic impurity~\cite{NB,Zaw}.
Such TLS may be formed accidentally in a strained glassy
material~\cite{AndHalpVarma}, or as a result of a
Jahn-Teller effect~\cite{MF2}, and TLS have also been
conjectured to occur at interfaces~\cite{Aliev}. Very
recently, the non-Fermi-liquid properties of the ground
state in this model have been invoked in the study of
dephasing rate of conduction electrons in disordered
metals~\cite{ZRvD}.

The degeneracy between the two states of the TLS based on
pseudospin symmetry is a crucial precondition for
two-channel Kondo screening and the formation of the
non-Fermi-liquid ground state. In practice this degeneracy
is almost always expected to be lifted because the
pseudospin, corresponding to parity about the center of the
TLS, is not in general a conserved quantity, and, in
particular, is not conserved by ordinary potential
scattering. This feature of the orbital two-channel Kondo
model has to be contrasted with magnetic Kondo models where
the relevant symmetry is time-reversal invariance, which is
broken only in special circumstances, {\em e.g.} by an
applied magnetic field or magnetic disorder. Therefore, an
analysis of the magnitude of disorder-induced splittings of
two-level systems is essential in evaluating the consistency
of the two-channel Kondo interpretation of the zero-bias
anomalies of Refs.~\onlinecite{RB,RB2}.

Most of the previous theoretical work on this
sub\-ject~\cite{WAM,MW} has been concentrated on computing
the second moment of the random splittings. A calculation of
$\left\langle\Delta^2\right\rangle$ induced by white-noise
potential scattering was reported in
Ref.~\onlinecite{WAM}. It was argued there that even small
amounts of disorder may lead to large splittings between the
energy levels of the TLS, thus effectively stopping the
Kondo screening at temperatures higher than $T_K$. However,
the distributions of splittings tend to be very asymmetrical
so that their moments are not representative of the typical
values. Moreover, the knowledge of the full distribution
function is necessary to understand how the splittings of an
ensemble of defects may affect the scaling behavior of
conductance.

It should be remarked that there is no direct evidence for
the existence of TLS in the nanoconstrictions studied in the
experiments of Refs.~\onlinecite{RB,RB2}. It is precisely
the match between the experimental scaling of conductance
and that predicted theoretically for two-channel Kondo
systems that is the main argument in favor of the
two-channel Kondo interpretation of the data.  The
theoretical scaling functions used for this purpose in
Refs.~\onlinecite{RB,RBLvD} were derived under the
assumption that no disorder other than the TLS themselves is
present, and thus it is of considerable interest to
understand how these scaling functions may be changed by
realistic amounts of static disorder.

In this paper we study distribution functions of splittings
for two models: (i) isotropic magnetic Kondo impurities in a
spin glass environment, and (ii) atomic TLS in an
environment of static defects inducing potential
scattering. In both cases the disorder is modeled by an
array of randomly located point scatterers. The magnetic
model may be realized if a dilute solution of weakly coupled
magnetic impurities undergoes Kondo screening in a
spin-glass environment, formed, for example, by a more
concentrated solution of more strongly coupled magnetic
impurities. The result for the distribution of splittings in
this case reproduces the distribution of internal fields in
spin glasses derived earlier in Ref.~\onlinecite{HK}. We
also analyze the effects of higher-order terms in the Kondo
coupling which cannot be reduced to RKKY-type
expressions. These terms are shown to lead to a finite
renormalization of the small-$\Delta$ portion of the
distribution functions, while their effect on the
large-$\Delta$ tail is negligible. A similar analysis is
performed for the distribution of splittings of atomic
TLS. The magnitude of the splittings obtained here should be
viewed as a lower bound, since only effects of electronic
disorder are taken into account, {\em i.e.}, we assume that
in the absence of such disorder the two states of the TLS
are degenerate~\cite{elastic}.

Using the distribution of splittings, we derive the
temperature dependence of the zero-bias conductance of a
metallic sample containing an ensemble of TLS. On this
basis, we re-analyze the two-channel Kondo interpretation of
the zero-bias anomalies~\cite{RB,RBLvD,RB2}. The discrepancy
between the observed scaling behavior and that derived in
the present work presents, in our view, a significant
challenge for the two-channel Kondo
interpretation. Furthermore, the estimate of the total
number of degenerate (in the absence of electronic disorder)
TLS which would be needed to produce the conductance
observed in Ref.~\onlinecite{RB} is found to be unphysically
large, indicating another problem with the two-channel Kondo
scenario.

The paper is organized as follows. In the next Section we
consider in greater detail the role of symmetry-breaking
disorder, present the main results, and discuss their
implications for the interpretation of the Ralph-Buhrman
experiments~\cite{RB}. Section~\ref{sec3} contains the
derivation of the splitting distribution for the magnetic
and non-magnetic Kondo effects. A brief discussion and the
summary are presented in Section~\ref{disc}.

\section{Disorder in the spin sector.}
\label{sec2}

The general $n$-channel anisotropic Kondo Hamiltonian for a
dynamic impurity $\hat{\bbox{\tau}}$ located at ${\bf r}=0$
has the form
\begin{eqnarray}
\label{nck}
H^{\left(n\right)} & = &\sum_{a=1}^n\sum_{\alpha} \int d{\bf
r}\psi_{\alpha a}^{\dagger}\left({\bf r}\right)
\left[\hat{\epsilon}\left(-i\nabla\right)\right]
\psi_{\alpha a}\left({\bf r}\right) \nonumber \\
& + &
\sum_{j=1}^3J_j\hat{\sigma}^j\left(0\right)\hat{\tau}_j,
\end{eqnarray}
where $\hat{\epsilon}$ is the Hamiltonian of band electrons,
$\hat{\sigma}^j\left({\bf r}\right)=\sum_{a=1}^n
\sum_{\alpha\beta} \psi_{\alpha a}^{\dagger}\left({\bf
r}\right)\sigma^j_{\alpha\beta} \psi_{\beta a}\left({\bf
r}\right)$ is the electron spin-density operator at ${\bf
r}$, $\bbox{\sigma}_{\alpha\beta}$ is the vector of Pauli
matrices, and $J_j$ are the exchange coupling constants. In
the isotropic case we use the notation $J=J_j$. Greek
indices are used to label spin quantum numbers, and Latin
indices denote channel quantum numbers~\cite{channels}.

\begin{figure}[tb]
\begin{center}
\leavevmode
\epsfxsize=3in
\epsfbox{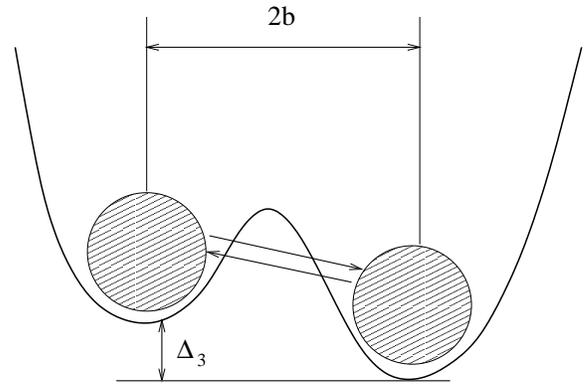}
\bigskip
\caption{\label{tls} Schematic representation of atomic
two-level system (TLS).}
\end{center}
\end{figure}

When considering the orbital two-channel realization of the
Kondo effect due to electron-TLS interaction, it is
convenient for our purposes to use, as a simple model of a
TLS, an atom which can tunnel between two minima of a
double-well potential located at ${\bf r}=\pm {\bf b}$
(Fig.~\ref{tls}).
The corresponding Hamiltonian has the form
\begin{eqnarray}
\label{eq18}
H & = & H_e+V_1\hat{\tau_1}\left(\psi^{\dagger}\left({\bf
b}\right)\psi\left(-{\bf b}\right)+\psi^{\dagger}
\left(-{\bf b}\right)\psi \left({\bf b}\right)\right) \\
\nonumber & + &
V_3\hat{\tau_3}\left(\psi^{\dagger}\left({\bf
b}\right)\psi\left({\bf b}\right)-\psi^{\dagger} \left(-{\bf
b}\right)\psi \left(-{\bf b}\right)\right),
\end{eqnarray}
where $H_e$ describes free electrons, and $\hat{\tau_1}$ and
$\hat{\tau_3}$ are Pauli matrices operating in the two-state
Hilbert space of the TLS. $V_{1}$ is the amplitude of the
``pseudospin-flip" process whereby the TLS tunnels between
its two states accompanied by the transfer of an electron
from ${\bf b}$ to $-{\bf b}$ or vice versa. $V_3$ is the
position-dependent interaction between the TLS and
electrons. The summation over physical spin is implied in
the terms bilinear in electronic annihilation and creation
operators $\psi$ and $\psi^{\dagger}$. Terms proportional to
$\hat{\tau_2}$ are absent in the Hamiltonian because of the
combined effect of the invariance under time reversal and
locality~\cite{Zaw}. Note that the apparently non-local term
proportional to $\hat{\tau_1}$ in Eq.\ (\ref{eq18}) is an
artifact of the approximation neglecting the full momentum
dependence of the coupling $V_1$.
\nopagebreak[4]

The Hamiltonian in Eq.\ (\ref{eq18}) can be cast in the form
of Eq.\ (\ref{nck}) with corresponding couplings~\cite{MF}
$J_3=\left(2\pi p_Fb/\sqrt{3}\right)V_3$ and $J_1=2\pi
V_1$. The channel quantum numbers in Eq.\ (\ref{nck}) would
then refer to physical spin, while spin quantum numbers
correspond to the impurity atom and electronic excitations
being located at either of the TLS potential minima, or,
depending on the choice of the basis, to different parity
eigenstates. Anisotropy of the couplings is unlikely to
occur in (hypothetic) magnetic realizations of this model,
while TLS realizations generically possess strong
anisotropy~\cite{Zaw}.

The disordered environment is modeled by adding to the
Hamiltonian in Eq.\ (\ref{nck}) the term
\begin{equation}
\label{disorder}
H_{\rm dis}=H_c+H_s=\int d{\bf r}\left[U_c\left({\bf
r}\right)\hat{n}\left({\bf r}\right)+{\bf U}_s\left({\bf
r}\right)\cdot\hat{\bbox{\sigma}} \left({\bf
r}\right)\right],
\end{equation}
where $\hat{n}$ and $\hat{\bbox{\sigma}}$ are the charge-
and spin-density operators of the conduction electrons, and
$U_c$ and ${\bf U}_s$ are the corresponding random
potentials.  Formally, the disorder Hamiltonian $H_{\rm
dis}$ introduces two additional energy scales into the
problem -- the inverse scattering time $\tau^{-1}$ and the
inverse spin-scattering time $\tau_s^{-1}$.

Before proceeding further, we will comment briefly on the
role of the charge disorder term $H_c$ in $H_{\rm dis}$. It
has been discussed in numerous works including
Refs.~\onlinecite{Fischer,ohkawa,Vladar,suga,FB,Kotliar}. A
related self-consistent model for the case of a {\em finite}
concentration of Kondo impurities has been considered in
Ref.~\onlinecite{Phillips}. This term affects only the
charge degree of freedom, and one of its effects is to
randomize the Kondo temperature $T_K$. In systems of lower
dimensionality, it can also produce singular corrections to
the energy dependence of thermodynamic and transport
coefficients in the perturbative high-temperature
regime. Incorporating charge disorder into the description
of the low-temperature ($T<T_K$) regime of the Kondo effect
has not been achieved so far. However, it has been argued in
Ref.~\onlinecite{ohkawa} that, in the one-channel magnetic
case, the basic nature of the ground state as a local Fermi
liquid would not change.

In the context of the magnetic Kondo effect, we will only
consider the role of spin disorder. Furthermore, we will
restrict the consideration to the case of the ordinary
one-channel effect, both because multichannel magnetic
realizations have not been unambiguously observed, and
because the consideration of the renormalization of
splittings (Section~\ref{sec3}) cannot be transferred to the
case $n >2$. [The renormalization of splittings in the $n=2$
case is discussed in Section~\ref{sec3}. The results for a
hypothetical magnetic isotropic $n=2$ case can be obtained
by a simple change $\gamma_m\rightarrow 2\gamma_m$ in Eq.\
(\ref{eq17}).]

In the orbital two-channel realization of the Kondo effect,
the roles of spin and orbital degrees of freedom are
partially interchanged. The potential scattering term $H_c$
in Eq.\ (\ref{disorder}) will produce a contribution
analogous to the spin-scattering term, $H_s$, when Eq.\
(\ref{eq18}) is transformed into the form of Eq.\
(\ref{nck})~\cite{note4}. In what follows, the term ``spin
scattering'' is understood to apply either to physical spin
in the context of magnetic Kondo effect, or to
``pseudospin'' constructed out of electronic states of
different parity in the context of the orbital two-channel
realization.

Our choice of model for disorder assumes that breaking of
the spin symmetry is due to a well-defined set of scatterers
present in the system. The scatterers couple to the spin
degrees of freedom, {\em e.g.} ``frozen'' magnetic
impurities in the magnetic one-channel case, or ordinary
non-magnetic defects in the orbital two-channel
case. Replacing such a set of defects by a continuous random
Gaussian-distributed potential, as is frequently done in
transport calculations, is not warranted here because the
distribution of splittings~$\Delta$ is non-universal. That
is, its form depends on the choice of the distribution
function for the random potentials $U_c\left({\bf r}\right)$
and ${\bf U}_s\left({\bf r}\right)$, and therefore a more
realistic model is required.

In the model of isolated scatterers spin disorder is given
by
\begin{equation}
\label{spindisorder}
{\bf U_s}\left({\bf r}\right)= g\sum_i{\bf
S}_i\delta\left({\bf r}-{\bf r}_i\right),
\end{equation}
where ${\bf S}_i$ are randomly oriented frozen spins located
at randomly selected points ${\bf r}_i$, and $g$ is the
corresponding exchange coupling constant. The distribution
of each of ${\bf S}_i$ is assumed isotropic and is given by
\begin{equation}
\label{eq2}
{\cal P}_S\left({\bf S}\right)=\frac{1}{2\pi}\delta
\left(1-{\bf S}^2\right).
\end{equation}

In the TLS case, we use a slightly more general expression,
allowing for scatterers of finite size:
\begin{equation}
\label{ranpot}
U_c\left({\bf r}\right)=\sum_iU\left({\bf r}-{\bf r}_i\right).
\end{equation}
Since ``spin'' degrees of freedom in this case are a subset
of orbital degrees of freedom, the above expression contains
both charge disorder and ``spin'' disorder terms. We will
only concentrate on the effect of the latter.

Below we will restrict our consideration to the case of
rotationally invariant potentials $U$, so that the
corresponding scattering matrix ${\cal T}$ can be reduced to
the diagonal form ${\cal T}={\mathrm{diag}}\left({\cal
T}_l\right)$, where ${\cal T}_l$ can be expressed in terms
of phase shifts $\eta_l$ for each value of the orbital
angular momentum as
\begin{equation}
\label{sctrmtx}
{\cal T}_l=-\frac{1}{\pi\nu}e^{i\eta_l}\sin\eta_l.
\end{equation}

The coordinates ${\bf r}_i$ of $N$ impurities are drawn from
a uniform distribution ${\cal P}_r\left(\left\{{\bf
r}_i\right\}\right)=1/{\cal V}^N$ where ${\cal V}$ is the
total volume of the sample. The calculations are performed
in the limit $N,{\cal V}\rightarrow\infty$ with the
concentration $c=N/{\cal V}$ kept finite. It is assumed that
the concentration is small in the sense that the typical
inter-impurity distance $d\sim c^{-1/3}$ is much larger than
the Fermi wavelength $\lambda_F=2\pi/p_F$, which implies, in
the magnetic case, $E_F\tau_s\gg 1$, and $E_F\tau\gg 1$ in
the TLS case, where $E_F$ and $p_F$ are the Fermi energy and
Fermi momentum, respectively. The effective ``spin''
scattering time in the TLS case is also of the order of
$\tau$, and we will keep the notation $\tau_s$ for it.

\subsection{Splitting of Internal States of Dynamic Impurity
by Spin Scattering.}

What is the main effect of spin scattering on the behavior
of a Kondo impurity? Even though a non-zero value of
$\tau_s^{-1}$ means that ``spin memory" of electrons has a
finite lifetime, the spin-flip processes at the location of
the Kondo impurity still lead to logarithmic divergences in
the high-temperature perturbative expansion (see
Appendix~\ref{suscept}).  In fact,~$\tau_s^{-1}$ does not
directly compete with the Kondo temperature~$T_K$.
Nevertheless, spin scattering can change the low-temperature
behavior of an impurity by introducing an effective
splitting,~$\Delta$, between its internal states.

To understand qualitatively why finite~$\tau_s^{-1}$ does
not by itself destroy the Kondo effect, consider the
underlying Anderson model for a magnetic impurity. In this
description, the Kondo effect is reflected in the
logarithmic divergence of the perturbative contributions to
the impurity electron self-energy at~$E_F$. There are two
processes that contribute to this self-energy: tunneling of
an impurity electron into the conduction band, and the
reverse process, tunneling of a conduction electron onto the
impurity. Both of these contributions are logarithmically
divergent at $E_F$ but with opposite signs. For an impurity
without on-site interactions, the two terms cancel and there
is no Kondo effect. Interactions remove this cancelation,
essentially because an occupied site can only decay in one
way -- by an electron tunneling out (double occupancy is
forbidden or strongly suppressed by Coulomb repulsion),
while an unoccupied site can decay in two ways -- by an
electron of either spin tunneling in. The appearance of a
finite $\tau_s^{-1}$ may change slightly the relative
tunneling rates for spin-up and spin-down electrons, but it
cannot significantly change the factor of two difference
between the rate of decay of an occupied and an unoccupied
site. Hence $\tau_s^{-1}$ does not directly destroy the
logarithmic divergences in perturbation theory associated
with the Kondo effect.
\begin{figure}[htb]
\begin{center}
\leavevmode
\epsfxsize=3in
\epsfbox{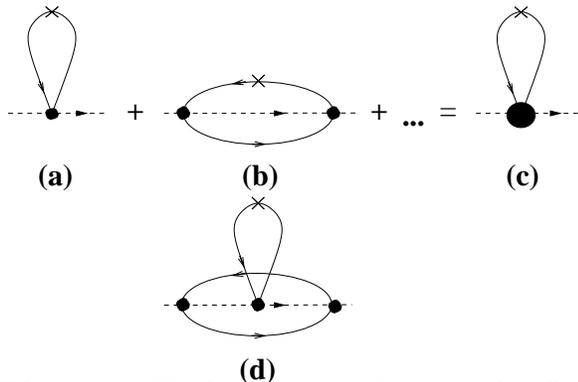}
\caption{\label{fig1} (a) The dominant contribution to the
effective splitting $\Delta$.  The solid line represents the
conduction electron Green function, and the broken line
represents pseudofermions. The dot represents the
interaction vertex $J$, and the cross corresponds to
impurity scattering. (b) The leading logarithmic order
(parquet) contribution to the splitting. (c) The sum of
parquet diagrams. The large dot represents the renormalized
(in parquet approximation) scattering amplitude $J_R$. (d)
Renormalization of splitting.}
\end{center}
\end{figure}

Importantly, however, the main effect of spin scattering is to break
time-reversal invariance and hence induce a splitting~$\Delta$ between
the two spin states of the Kondo impurity. The splitting results from
the appearance of a random non-zero quantum-mechanical expectation
value of the local spin density of states
$\bbox{\sigma}\left(\epsilon\right)$ at the impurity site. The
coupling of $\bbox{\sigma}\left(\epsilon\right)$ to the dynamic
impurity via~$J$ is responsible for breaking the energy degeneracy
between different orientations of the impurity spin. In the
diagrammatic language, the splitting is associated {\em not} with the
modification of the standard set of logarithmic diagrams, but rather
with proliferation of a new set of diagrams which were forbidden by
${\rm SU}(2)$ and time-reversal symmetries in the absence of spin
disorder. The leading contribution to~$\Delta$ in perturbation theory
in the coupling strength $J$ is shown in Fig.~\ref{fig1}a. The energy
scale established by~$\Delta$ serves as a cut-off of all logarithmic
divergences in the perturbation theory in~$J$.

Formally, the main effect of spin disorder is to generate a
self-energy term which is essentially energy-independent,
and can be reinterpreted as a contribution to the effective
impurity Hamiltonian. It is a Hermitian $2\times 2$ matrix
in the space of impurity states. In the magnetic case, it is
convenient to expand this matrix in the basis of Pauli
matrices as
\begin{equation}
\label{eq3}
H_{\mbox{\text eff}}=\sum_{\alpha
=1}^3\Delta_{\alpha}\hat{\tau}^{\alpha},
\end{equation}
where the~$\Delta_{\alpha}$ denote the components of the
impurity energy in this basis.

The discussion presented so far does not distinguish between
the one- and two-channel cases. This is natural since the
high-temperature diagrammatic expansions in the two cases
have identical structures, and differ only in factors of $2$
(from the two channels) for closed electronic loops. If the
Kondo temperature $T_K < \Delta$, spin-flip processes are
suppressed at $T<\Delta$, the strongly correlated state is
never formed, and below~$\Delta$ all temperature dependences
are of the Fermi liquid type. If, however, $T_K > \Delta$,
the behavior of the one- and two-channel Kondo systems is
very different. In the one-channel case a non-zero but small
$\Delta$ is equivalent to a weak polarizing field acting on
the Kondo impurity, resulting in finite but small changes in
the values of impurity susceptibility and
conductance. However, since these quantities depend on
$T_K$, which is itself altered in a random way by disorder,
no significant experimental consequences follow from a small
splitting $\Delta < T_K$.

Conversely, in the two-channel case when~$\Delta \ne 0$
there appear two distinct low-temperature regimes. A
non-Fermi-liquid regime survives in the interval between
$T^*<T<T_K$, where $T^*$ is a new characteristic
temperature, $T^*=\Delta^2/2\pi
T_K$~\cite{PangCox,SacraSchlott,sengupta}. A Fermi-liquid
regime emerges below $T^*$. Hence, the low-temperature
properties of an ensemble of two-channel Kondo impurities
will explicitly depend on the distribution of
splittings~$\Delta$.

If $\Delta >T_K$, a non-Fermi-liquid regime does not exist,
and at temperatures below~$\Delta$ the splitting becomes the
only relevant energy scale. In other words, the TLS (or,
rather, the composite object comprised of the TLS and
correlated electrons) becomes frozen in its lowest energy
state. Excitations above this state result in $T^2$
dependences which blend with other Fermi-liquid effects that
are always present. As was demonstrated by Moustakas and
Fisher~\cite{MF}, a generic set of TLS parameters almost
always corresponds to $\Delta > T_K$. In order to observe
the non-Fermi-liquid square-root scaling behavior, the
parameters of the Hamiltonian have to be fine tuned~\cite{remark}.

The distribution of splittings in the ordinary (one-channel)
magnetic case has a simple form (Ref.~\onlinecite{HK},
Section~\ref{sec3})
\begin{equation}
\label{eq17}
{\cal P}_m\left(\Delta\right) = \frac{4\gamma_m}{\pi}
\frac{\Delta^2}{\left(\Delta^2+\gamma_m^2\right)^2},
\end{equation}
where $\gamma_m=\left(2\pi^2 c/3p_F^3\right)Jg\nu^2E_F$ is a
constant which determines the scale for the typical values
of~$\Delta$.  It is proportional to the strength of the
dimensionless Kondo coupling $\nu J$, where $\nu$ is the
density of states at the Fermi level, and to the magnitude
of spin scattering $g$ [Eq.\ (\ref{spindisorder})]. The
quadratic suppression of ${\cal P}_m$ at small~$\Delta$
results from the fact that~$\Delta^2$ is a sum of three
random variables~$\Delta_{\alpha}^2$, each possessing a
smooth distribution near $\Delta_{\alpha}=0$. Crucially, the
Kondo temperature $T_K\sim E_Fe^{-1/\left(\nu J\right)}$ is
exponentially small, so even in systems with weak magnetic
disorder ($c/p_F^3,\,\nu g\ll 1$) $\gamma_m$ may be
comparable to $T_K$.

In the orbital two-channel Kondo case, we will show in
Section~\ref{sec3} that ${\cal P}\left(\Delta\right)$ is
described by a more complicated analytical expression,
\begin{equation}
\label{eq23}
{\cal P}\left(\Delta\right)=\int_0^{2\pi}
\frac{d\theta}{2\pi}\frac{K
\left(\theta\right)}{\sqrt{K_1K_3}}\frac{\Delta\gamma
\left(\theta\right)}
{\left[\Delta^2+\gamma^2\left(\theta\right)\right]^{3/2}},
\end{equation}
where $K\left(\theta\right)=\left(\cos^2\theta/K_1
+\sin^2\theta/K_3\right)^{-1}$, $\gamma\left(
\theta\right)=\displaystyle{\frac{2\pi c}{3p_F^3}
\sqrt{\frac{K\left(\theta\right)}{2}}F\left(
\theta\right)}$, and
\begin{equation}
\label{eq24}
F\left(\theta\right)=1+\frac{\cos^2\theta}{\left|\sin\theta
\right|}\ln\left(\left|\tan\theta\right|+
\frac{1}{\left|\cos\theta\right|}\right).
\end{equation}
The constants $K_1$ and $K_3$ depend on the strength of
electronic coupling to the TLS and static scatterers,
\begin{equation}
\label{Ks}
K_1=2\left(E_F V_1\nu t\right)^2,\,\,\,
K_3=2\left(E_F V_3\nu t\right)^2\left(2p_Fb\right)^2,
\end{equation}
where the scatterer strength $t$ is expressed in terms of
the scattering phase shifts $\eta_l$ introduced in Eq.\
(\ref{sctrmtx}):
\begin{equation}
\label{t}
t=\sum_l\left(-1\right)^l\left(2l+1\right)
\sin\left(2\eta_l\right).
\end{equation}
The function $\gamma\left(\theta\right)$ is the analog of
the parameter $\gamma_m$ introduced in the magnetic case.

In the asymptotic limit
$\Delta\gg\displaystyle{\frac{c}{p_F^3}}
\max\left\{\sqrt{K_1},\sqrt{K_3}\right\}$ the above
expression simplifies to
\begin{equation}
\label{eq25}
{\cal P}\left(\Delta\right)\sim
\frac{\sqrt{2}cK_1K_3}{3p_F^3\Delta^2}\int_0^{2\pi}
\frac{F\left(\theta\right)d\theta}{\left(
K_1\sin^2\theta+K_3\cos^2\theta\right)^{3/2}}.
\end{equation}

The distribution function in Eq.\ (\ref{eq25}) has the same
asymptotic $\Delta^{-2}$ behavior as Eq.\ (\ref{eq17}) for
large~$\Delta$. However, the full distribution function
given by Eq.\ (\ref{eq23}) is linear rather than quadratic
at small~$\Delta$ because, in the absence of $\hat{\tau_2}$
terms in the bare Hamiltonian,~$\Delta^2$ is a sum of two
rather than three random variables. In physical models of
TLS the couplings are usually related via $V_1 \sim V_3
\left(2p_Fb\right)^2$ (see Ref.~\onlinecite{Zaw}). This 
typically corresponds to $K_3 \gg K_1$, so that the
distribution function acquires a third asymptotic region. It
extends between the sharp maximum at $\Delta\sim
\left(c/p_F^3\right)\sqrt{K_1} \sim \left(\nu
V_1ct/p_F^3\right)E_F$ and the beginning of 
the~$\Delta^{-2}$ decay at $\Delta\sim
\left(c/p_F^3\right)\sqrt{K_3} \sim \left(\nu
V_3ct/p_F^3\right)E_F$, and corresponds to approximately
linear decay of the distribution function from a constant
value. If $K_1\approx K_3$, the intermediate asymptotic
regime disappears.

\begin{figure}[ht]
\begin{center}
\leavevmode
\epsfxsize=3in
\epsfbox{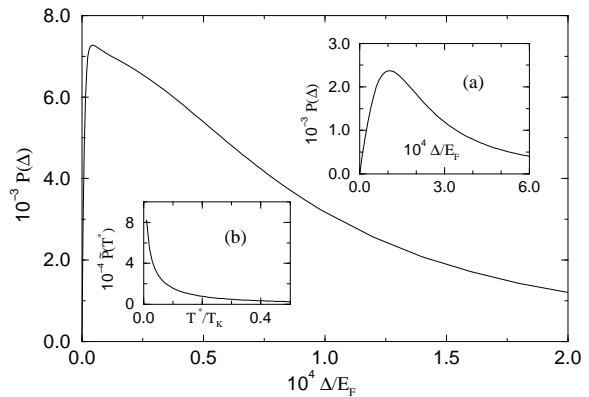}
\caption{\label{pofdelta} The distribution function of
splittings $\Delta$. The parameters of the distribution
function are chosen as follows(see Appendix~\protect\ref{param}):
$c/p_F^3=10^{-4}$, $K_3=.38 E_F^2$, $K_1=3.7\cdot
10^{-5}E_F^2$; the Fermi energy of Cu is $E_F=8\cdot
10^4$K. Insets: (a) Isotropic case $K_1=K_3=.5 E_F^2$. (b)
Distribution of the crossover temperature $T^*$; parameter
regime is the same as in the main graph; $T_K\approx
8.2{\mathrm K}$.}
\end{center}
\end{figure}

The graph of ${\cal P}\left(\Delta\right)$ for the strongly
anisotropic regime, corresponding to the choice of
parameters discussed in Appendix~\ref{param}
($c/p_F^3\approx 10^{-4}$, $K_3=.38 E_F^2$, $K_1=3.7\cdot
10^{-5}E_F^2$), is shown in Fig.~\ref{pofdelta}. The
isotropic case ($K_1=K_3=.5 E_F^2$) is shown in the top
inset. Note that both crossovers between asymptotic regimes
in the main graph occur at values of~$\Delta$ which are
significantly smaller than the r.m.s. splitting quoted in
Ref.~\onlinecite{WAM} (see also Ref.~\onlinecite{kozub}):
$\left\langle\Delta^2\right\rangle^{1/2}\sim
100\mathrm{K}\sim 10^{-3}E_F$, where the value $E_F\approx
8\cdot 10^4K$ for copper has been used.

Assuming that fluctuations of the Kondo temperature $T_K$
can be neglected (see Appendix~\ref{randomtk}), the
distribution of $T^*$ -- the temperature at which the
crossover from non-Fermi-liquid to Fermi-liquid behavior
occurs -- is easily inferred from Eq.\ (\ref{eq23}):
\begin{equation}
\label{w}
\widetilde{\cal P}\left(T^*\right)=\int_0^{2\pi}
\frac{d\theta}{4\pi}\frac{K
\left(\theta\right)}{\sqrt{K_1K_3}}\frac{\gamma
\left(\theta\right)/\sqrt{2\pi T_K}}
{\left[T^*+\gamma^2\left(\theta\right)/2\pi
T_K\right]^{3/2}}.
\end{equation}
This expression is applicable only for $T^*\ll T_K$, where
the relationship $T^*=\Delta^2/2\pi T_K$ holds. The graph of
$\widetilde{\cal P}\left(T^*\right)$ is shown in the bottom
inset in Fig.~\ref{pofdelta}. The corresponding limiting
behaviors are
\begin{equation}
\label{mainres}
\widetilde{\cal P}\left(T^*\right) \propto \left\{ \begin{array}
{l}
\displaystyle{\frac{4T_K}{\sqrt{K_1K_3}}\left(\frac{2\pi
c}{3p_F^3}\right)^{-2}\int_0^{\pi/2}
\frac{d\theta}{F^2\left(\theta\right)}},  \\ 
\quad\quad\quad\quad\quad T^* <
\left(c/p_F^3\right)^2{\mathrm{min}}\left(K_1,K_3\right)/T_K,
\\ \displaystyle{\frac{2\pi c/3p_F^3}{\sqrt{\pi^3T_KT^{*3}}}
\int_0^{\pi/2}d\theta\frac{K^{3/2}\left(\theta\right)
F\left(\theta\right)}{\sqrt{K_1K_3}}},  \\
\quad\quad\left(c/p_F^3\right)^2{\mathrm{max}}
\left(K_1,K_3\right)/T_K <T^* <T_K,
\end{array} \right.
\end{equation}
In the intermediate regime
$\left(c/p_F^3\right)^2K_1/T_K<T^*<
\left(c/p_F^3\right)^2K_3/T_K$
the distribution function $\widetilde{\cal
P}\left(T^*\right)$ is proportional to $1/\sqrt{T^*}$. Both
the intermediate and the far asymptotic regimes may not
exist if the corresponding boundaries become comparable to,
or larger than, $T_K$.

\subsection{Relevance of Disorder to Two-Channel-Kondo-Model
Interpretation of Zero-Bias Anomalies.}
\label{experiment}

If the parameters of TLS are randomly drawn from a certain
distribution, the net contribution to conductance $\Delta
G_N\left(T\right)$ from $N$ TLS can be approximated as
\[\Delta G_N\left(T\right)\approx N\left\langle 
G\right\rangle + O\left(\sqrt{N}\right),\] where
$\left\langle G\right\rangle$ is the contribution of a
single TLS averaged over the distribution of $T^*$. It was
estimated in Ref.~\onlinecite{RB} that at least 10 separate
TLS in the vicinity of the nanoconstriction, and up to 40 in
some samples, must be present in order to explain the
observed magnitude of the zero-bias anomaly. In this regime
the second term on the right-hand side in the above
expression would manifest itself as noise on the
experimental curves, and hence can be neglected. A scaling
ansatz for the anomaly in the zero-bias conductance due to a
single TLS can be represented at $T<T_K$ as $\Delta
G(T)=CT^{1/2}\Upsilon(T^*/T)$, where $C$ is a constant and
$\Upsilon$ is a smooth function with the limiting behavior
$\Upsilon(0)=1$ and $\Upsilon(x\gg 1)\rightarrow 0$. The
signal from $N$ TLS is then written as
\begin{mathletters}
\begin{eqnarray}
\label{eq30a}
\Delta G_N\left(T\right) & = & NCT^{1/2}\int\widetilde{\cal P}
\left(T^*\right)\Upsilon\left(T^*/T\right)dT^* \nonumber \\
& = & NCT^{3/2}
\int\widetilde{\cal
P}\left(xT\right)\Upsilon\left(x\right)dx,
\end{eqnarray}
where $x=T^*/T$, and $\widetilde{\cal P}\left(T^*\right)$ is
a normalized distribution. The function
$\Upsilon\left(x\right)$ reflects ``freezing out'' of some
impurity degree of freedom, and is likely to decay
exponentially at large $x$. Thus the convergence of the
above integral at large $x$ is provided by either $\Upsilon$
or $\widetilde{\cal P}$ depending on the temperature.

It has been conjectured in Ref.~\onlinecite{vonDelft&Co}
that the observed $T^{1/2}$ scaling of $G_N$ may be
consistent with the existence of a distribution of
splittings because of an auto-selection process, in which
only impurities with sufficiently small~$\Delta$ contribute
to conductance. Such a scenario would imply that the
integral in Eq.\ (\ref{eq30a}) could be approximated as
\begin{equation}
\label{eq30b}
\Delta G_N\left(T\right)\approx 
NCT^{3/2}\Upsilon\left(0\right)\int\widetilde{\cal
P}\left(xT\right)dx,
\end{equation}
\end{mathletters}
resulting in $\Delta G_N=NCT^{1/2}$ by virtue of the
normalization of $\widetilde{\cal P}$. However, this
approximation is only valid if $\widetilde{\cal
P}\left(T^*\right)$ is peaked so sharply that the integral
in Eq. (\ref{eq30a}) is not much different from its value in
the limiting case $\widetilde{\cal
P}\left(T^*\right)\propto\delta\left(T^*\right)$. We will
now consider under what circumstances such a behavior of
$\widetilde{\cal P}\left(T^*\right)$ is possible, and what
consequences a different behavior of $\widetilde{\cal
P}\left(T^*\right)$ would have.

Randomly formed TLS in metallic glasses typically have a
broad distribution of asymmetries~$\Delta_z$ even in the
absence of electronic disorder~\cite{Black}. Neglecting
contributions from random~$\Delta_x$ at first, we can write
${\cal P}\left(\Delta\right) = 1/W_z$, where $W_z$ is
independent of~$\Delta$ in the region of interest. The
corresponding distribution of $T^*$ is $\widetilde{\cal
P}\left(T^*\right)=\sqrt{\pi T_K/2T^*}/W_z$, and
consequently $\Delta G_N$ must display a linear in
temperature behavior,
\[
\Delta G_N=NC\sqrt{\frac{\pi T_K}{2W_z^2}}T
\int_0^{\infty}\frac{dx}{\sqrt{x}}\Upsilon\left(x\right).
\]
Non-zero values of~$\Delta_x$ distributed with a width $W_x
<W_z$ would only exacerbate the discrepancy with the
experimentally observed $\Delta G_N\sim T^{1/2}$ behavior:
specifically, the distribution of~$\Delta$ would acquire a
linear dip at $\Delta < W_x$, yielding a flat distribution
of $T^*$, leading in turn to $\Delta G_N\left(T\right)\sim
T^{3/2}$ from the integral in Eq.\ (\ref{eq30a}). Thus in
order for the Kondo effect in its orbital two-channel
realization to be the cause of the observed $T^{1/2}$
scaling, one must assume a set of nearly degenerate TLS, at
least before disorder is taken into account. Glassiness as a
source of TLS in Ralph-Buhrman samples must, therefore, be
ruled out.

\begin{figure}[ht]
\begin{center}
\leavevmode
\epsfxsize=3in
\epsfbox{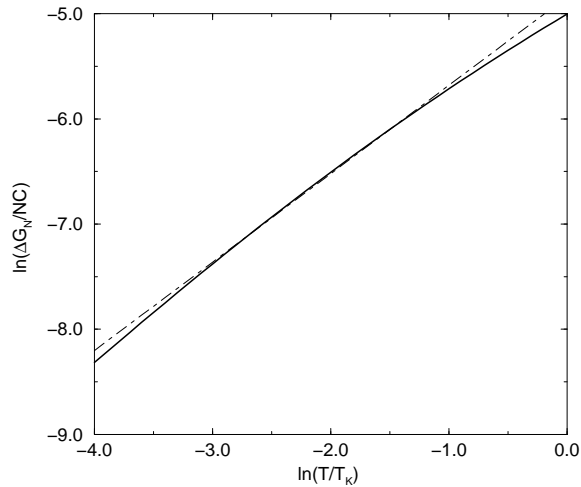}
\caption{\label{condplot} Temperature dependence of the
change in conductance due to scattering by an ensemble of
TLS, based on Eq.\ (\protect\ref{condres}). The solid line is the
result of numerical evaluation, and the dot-dash line is the
linear fit $\ln\left(\Delta
G_N/NC\right)=-4.83+0.84\ln\left(T/T_K\right)$.}
\end{center}
\end{figure}

Let us now turn to the case when electronic disorder is the
only source of TLS splitting, i.e. the TLS are assumed to be
formed by some mechanism which, in the absence of coupling
to conduction electrons, ensures their degeneracy. The
analytic form for the distribution of $T^*$, Eq.\ (\ref{w}),
can be substituted in Eq.\ (\ref{eq30a}) together with an
exponential ansatz $\Upsilon\left(x\right)=e^{-x}$. Using
the available experimental data to determine the parameters
of the distribution $\widetilde{\cal P}\left(T^*\right)$ in
Eq.\ (\ref{w}) is not straightforward, and is discussed in
detail in Appendix~\ref{param}. Integrating over $x$ we
obtain the following integral representation for the TLS
contribution to conductance:
\begin{eqnarray}
\label{condres}
 & & \Delta G_N\left(T\right)=NCT^{1/2}
\int_0^{2\pi}\frac{d\theta}{2\pi}
\frac{K\left(\theta\right)}{\sqrt{K_1K_3}} \nonumber \\
& & \times\left\{1-
\frac{\gamma\left(\theta\right)}{\sqrt{2T_KT}}
\exp\left[\frac{\gamma^2\left(\theta\right)}{2\pi
T_KT}\right]{\mathrm{erfc}}
\left[\frac{\gamma\left(\theta\right)}{\sqrt{2\pi
T_KT}}\right]\right\}.
\end{eqnarray}
The remaining integral over $\theta$ is performed
numerically, and the resulting graph for the temperature
dependence of $\Delta G_N$ is shown in Fig.~\ref{condplot}.
As discussed in Appendix~\ref{param}, electronic disorder is
assumed to be caused by strongly scattering non-equilibrium
vacancies. The plot in Fig.~\ref{condplot} does indicate a
behavior close to a power law. However, the best fit in the
region between $.05T_K<T<.5T_K$, corresponding to the
interval $.4K<T<4K$ for $T_K\approx 8K$, gives $\Delta
G_N\propto T^{\alpha_{\mathrm{fit}}}$ with
$\alpha_{\mathrm{fit}}\approx .84$. The deviation from the
$T^{1/2}$ behavior expected for degenerate TLS is rather
significant. While the measured exponent in the temperature
dependence of zero-bias conductance does show deviations
from $\alpha=1/2$, they do not exceed $.25$
(Ref.~\onlinecite{vonDelft&Co}), so that the value $.84$
clearly lies outside the experimental error.

This result can be understood better by examining under what
conditions Eq.\ (\ref{eq30b}) can be valid. Let us assume,
for simplicity, that $\widetilde{\cal P}$ is characterized
by a single parameter, its width $W$. Formally, Eq.\
(\ref{eq30b}) can be used when $T\ll W$. Let us consider a
hypothetical case of a sufficiently sharp distribution,
e.g. a Gaussian, $\widetilde{\cal P}=\sqrt{2/\pi
W^2}\exp\left[-T^{*2}/2W^2\right]$. Substituting this form
into Eq.\ (\ref{eq30a}) we find that in the temperature
region $W<T<10 W$, the best fit gives a power-law exponent
of approximately .7, while the asymptotic behavior .5 is
approached within a 10$\%$ accuracy in the region $10 W<T<20
W$. Thus, even in the case of a sharp distribution, the
relation $T\ll W$ has to be understood as implying at least
an order-of-magnitude difference. Returning to the actual
distribution Eq.\ (\ref{eq23}), we note that it is a much
broader function with a power-law $1/T^{*3/2}$ tail so that
the condition for the validity of Eq.\ (\ref{eq30b}) is even
stricter. At the same time, the larger of the two parameters
controlling the width of $\widetilde{\cal
P}\left(T^*\right)$ is $W\sim
\left(c/p_F^3\right)^2 K_3/T_K\sim 3{\mathrm K} \sim .3
T_K$, so that the asymptotic condition $W\ll T$ can never be
satisfied for temperatures below $T_K$.

It should be observed that the most optimistic choice of
parameters for the two-channel Kondo interpretation,
corresponding to scattering by non-equilibrium vacancies,
leads to an estimate for the concentration of defects
$c/p_F^3\approx 10^{-4}$.  Using such a choice of parameters
to support the two-channel Kondo interpretation also meets
with a difficulty concerning the fraction of defects which
form two-level systems. Indeed, using the estimate made in
Ref.~\onlinecite{RB} of up to 40 separate TLS in the
vicinity of the nanoconstriction, the corresponding estimate
for the density of {\em active} TLS is found in
Ref.~\onlinecite{vonDelft&Co} to be $10^{-4}/\mathrm{atom}$,
or $c_{\mathrm{active}}/p_F^3\approx 2\cdot 10^{-5}$. It can
be assumed that TLS with $T^*< T_0$, where $T_0$ is the
appropriately chosen cut-off temperature, are active. Then
the ratio of the concentration of active TLS to their total
concentration $c_{\mathrm{TLS}}$ is given by
\begin{equation}
\label{fraction}
\frac{c_{\mathrm{active}}}{c_{\mathrm{TLS}}}=
\int_0^{T_0}\widetilde{\cal P}\left(T^*\right)dT^*.
\end{equation}
Choosing $T_0\approx .8K$, which is a slightly generous
assumption, since the $T^{1/2}$ behavior is traced
experimentally down to temperatures $T\approx .4K$, we find
$c_{\mathrm{active}}/c_{\mathrm{TLS}}\approx 0.4$. The total
density of TLS is consequently estimated at
$c_{\mathrm{TLS}}/p_F^3\approx
\left(1/0.4\right)c_{\mathrm{active}}/p_F^3\approx 0.5 \cdot
10^{-4}$. Comparing this to the above estimate of
$c/p_F^3\approx 10^{-4}$ for the total concentration of
defects forces the improbable conclusion that {\em half} of
all the defects in the constriction are two-level
systems. In other words, in order for the auto-selection
mechanism to work, the total number of TLS which would be
degenerate in the absence of disorder must be so large as to
be inconsistent with the results of measurements indicating
a rather small overall density of defects in the
nanoconstriction. Although indirect, this reasoning serves,
in our view, as another indicator of internal consistency
problems with the two-channel Kondo interpretation of
zero-bias anomalies in Cu nanoconstrictions.

\section{The distribution of splittings.}
\label{sec3}

\subsection{Local Spin in a Spin Glass Environment}

The coupling of the conduction electrons to the random spins
results in the appearance of a non-zero expectation value of
the conduction-electron spin density at a generic point in
the sample. When this spin density is coupled to the dynamic
impurity, the lowest order effect is to generate a
self-energy matrix, whose components~$\Delta_{\alpha}$ in
the basis of Pauli matrices and to the lowest order in
impurity concentration $c$ are (see diagram in
Fig.~\ref{fig1}a)
\begin{equation}
\label{eq4}
\Delta_{\alpha}=-iJg\sum_i\int\frac{d\epsilon}{2\pi}G^2
\left(\epsilon;{\bf r}_i\right)S_i^{\alpha},
\end{equation}
where the zero-temperature Green function in the coordinate
representation is given by
\begin{equation}
\label{eq5}
G\left(\epsilon;{\bf
r}\right)=-\frac{\pi\nu}{p_Fr}\exp\left\{i\left(p_Fr+ 
\frac{\epsilon}{v_F}r\right)\mbox{\text Sgn}\epsilon\right\}.
\end{equation}
Note that~$\Delta_{\alpha}$ is proportional to the
RKKY-induced random spin polarization in direction $\alpha$
at the position of the impurity. To the lowest order in $J$,
the distribution of splittings follows the distribution of
internal random magnetic fields in spin glasses. The latter
has been derived in Ref.~\onlinecite{HK} using a somewhat
simplified RKKY interaction, in which its oscillatory
character is modeled by random signs. The derivation below,
while reproducing the essential results of
Ref.~\onlinecite{HK} serves primarily to introduce the more
technically involved derivation for the non-magnetic
two-channel case presented in the next subsection.

Integrating over energy we obtain
\begin{equation}
\label{eq6}
\Delta_{\alpha}=Jg\nu^2\sum_i
\frac{\pi E_F}{\left(p_Fr_i\right)^3}
\cos\left(2p_Fr_i\right)S_i^{\alpha}.
\end{equation}
The distribution function of $\Delta=\sqrt{\sum_{\alpha}
\Delta_{\alpha}^2}$ therefore takes the form
\begin{eqnarray}
\label{eq7}
& & {\cal P}_m\left(\Delta\right) = 2\Delta\int\prod_i^N\frac{d{\bf
r}_i}{\cal V} \int\prod_i^N \frac{d{\bf
S}_i}{2\pi} \nonumber \\
& \times & \delta\left(1-{\bf
S}_i^2\right)\delta\left(\Delta^2-\frac{K}{2} \sum_{ij}
f\left({\bf r}_i\right)f\left({\bf r}_j\right){\bf S}_i {\bf
S}_j\right),
\end{eqnarray}
where $K=2\left(\pi Jg\nu^2E_F\right)^2$ and $f\left({\bf
r}_i\right)\equiv f_i= \displaystyle{\frac{\cos\left(2p_Fr_i
\right)}{\left(p_Fr_i\right)^3}}$. Exponentiating the second
$\delta$-function and introducing a shorthand notation ${\bf
F}=\sum_if_i{\bf S}_i$ we rewrite the distribution function
as
\begin{eqnarray}
\label{eq8}
{\cal P}_m\left(\Delta\right)& = & 2\Delta\int\prod_i^N\frac{d{\bf
r}_i}{\cal V} \int\prod_i^N \frac{d{\bf
S}_i}{2\pi}\delta\left(1-{\bf S}_i^2\right) \nonumber \\
&\times &\int\frac{d\mu}{2\pi}e^{i\mu\Delta^2-\frac{i}{2}\mu K{\bf
F}^2}.
\end{eqnarray}
Decoupling the last term in the exponent with the help of a
``Hubbard-Stratonovich" transformation we obtain
\begin{eqnarray}
\label{eq9}
{\cal P}_m\left(\Delta\right) & = &
2\Delta\int\prod_i^N\frac{d{\bf r}_i}{\cal V} \frac{d{\bf
S}_i}{2\pi}\delta\left(1-{\bf S}_i^2\right)
\int\frac{d\mu}{2\pi}e^{i\mu\Delta^2} \nonumber \\
&\times & \int\frac{d\bbox{\lambda}}
{\left(2\pi i\mu K\right)^{3/2}}
\exp\left\{\frac{i\bbox{\lambda}^2}{2\mu K}
-i\bbox{\lambda}{\bf F}\right\} \nonumber \\ & = &
2\Delta\int\frac{d\mu}{2\pi}e^{i\mu\Delta^2}
\int\frac{d\bbox{\lambda}}{\left(2\pi i\mu K\right)^{3/2}}
\exp\left\{\frac{i\bbox{\lambda}^2}{2\mu K}\right\}
\nonumber \\
& \times & \left[\int\frac{d{\bf r}}{\cal V}
\frac{d{\bf S}}{2\pi}\delta
\left(1-{\bf S}^2\right)e^{-i f\left({\bf r}
\right)\bbox{\lambda}{\bf S}}\right]^N.
\end{eqnarray}
The expression in the square brackets can be transformed as
follows:
\begin{eqnarray}
\label{eq10}
& & \left[\int\frac{d{\bf r}}{\cal V}\int\frac{d{\bf
S}}{2\pi}\delta \left(1-{\bf S}^2\right)\exp\left\{-i
f\left({\bf r} \right)\bbox{\lambda}{\bf S}
\right\}\right]^N \nonumber \\
& = & \left[\int\frac{d{\bf r}}{\cal V}\frac{\sin\left[
\left|\bbox{\lambda}\right|f\left({\bf r}\right)
\right]}{\left|\bbox{\lambda}\right|f\left({\bf r}
\right)}\right]^N \nonumber \\ 
& \approx & \exp\left\{-c\int
d{\bf r}\left(1-
\frac{\sin\left[\left|\bbox{\lambda}\right|f \left({\bf
r}\right)\right]}{\left|\bbox{\lambda} \right|f\left({\bf
r}\right)}\right)\right\}.
\end{eqnarray}
To compute the last integral the following approximation is
employed: since the dominant contribution to the
distribution function is expected to come from $r\sim d$, it
is possible to decouple fast oscillations in $f$
(proportional to $\cos 2p_Fr$) from the slow decay
$\left(p_Fr\right)^{-3}$. Formally, $f\left({\bf r}\right)$
is replaced with $f\left(\varphi,{\bf
r}\right)=\displaystyle{
\frac{\cos\varphi}{\left(p_Fr\right)^3}}$ with a
simultaneous replacement $\displaystyle{\int d{\bf
r}\rightarrow \int d{\bf
r}\int\frac{d\varphi}{2\pi}}$. Integration over ${\bf r}$
now gives
\begin{equation}
\label{eq11}
\exp\left\{-\int\frac{d\varphi}{2\pi}
\frac{\pi^2c\left|\bbox{\lambda}\right|\left|\cos\varphi
\right|}{3p_F^3}\right\}=\exp\left\{-\frac{2\pi c\left|
\bbox{\lambda}\right|}{3p_F^3}\right\}.
\end{equation}

Absorbing $\sqrt{|\mu|K}$ into $\lambda$ we can rewrite the
distribution function as
\begin{eqnarray}
\label{eq12}
& & {\cal P}_m\left(\Delta\right)=
2\Delta\int\frac{d\mu}{2\pi}e^{i\mu\Delta^2}
\frac{4\pi}{\left(2\pi
i{\mathrm{Sgn}}\mu\right)^{3/2}} \nonumber \\
&\times & \int_0^{\infty}
\lambda^2d\lambda\exp\left\{\frac{i{\mathrm{Sgn}}\mu}{2}
\lambda^2-\frac{2\pi c\sqrt{\left|\mu\right|K}
\lambda}{3p_F^3}\right\}.
\end{eqnarray}
Rotating the contour of integration by
$\left(\pi/4\right){\mathrm{Sgn}}\mu$ we arrive at
\begin{eqnarray}
\label{eq13}
{\cal P}_m\left(\Delta\right)=\int
&\displaystyle{\frac{d\mu}{2\pi}} & e^{i\mu\Delta^2} 
\displaystyle{\sqrt{\frac{2}{\pi}}} 
\int x^2dx\exp \left\{-\frac{1}{2}x^2 \right. \nonumber \\
& - & \left. \frac{2\pi
c}{3p_F^3}\frac{1+i{\mathrm{Sgn}}\mu}{\sqrt{2}}
\sqrt{\left|\mu\right|K}x\right\}.
\end{eqnarray}

Performing another change of variables
$y^2=2\left|\mu\right|\Delta^2$ and rotating the contour of
integration over $y$ we obtain the following integral
representation for the distribution function:
\begin{eqnarray}
\label{eq14}
{\cal P}_m\left(\Delta\right) & = &
-\left(\frac{2}{\pi}\right)^{3/2}
{\mathrm{Re}}\frac{\partial}{\partial\gamma_m}
\int_0^{\infty}xdx\int_0^{\infty}dy \nonumber \\
& \times & \exp\left\{-\frac{1}{2}
\left(x^2+y^2\right)+i\frac{\gamma_m}{\Delta} xy\right\},
\end{eqnarray}
where $\gamma_m=\displaystyle{\frac{2\pi c}{3p_F^3}
\sqrt{\frac{K}{2}}}$. Using polar coordinates
$\left(\rho,\theta\right)$ in the $\left(x,y\right)$ plane
and integrating over $\rho$ the distribution function can be
rewritten as
\begin{equation}
\label{eq15}
{\cal P}_m\left(\Delta\right)=-\frac{2}{\pi} {\rm
Re}\frac{\partial}{\partial\gamma_m} \int_0^{\pi/2}
\frac{\cos\theta
d\theta}{\left(1-i\displaystyle{\frac{\gamma_m}{\Delta}}
\sin 2\theta\right)^{3/2}}.
\end{equation}
The remaining integral can be performed by elementary means
leading to the expression for the distribution function
(cf. Ref.~\onlinecite{HK}) quoted in Eq.\ (\ref{eq17}).

\subsection{Non-Magnetic TLS in the Presence of 
Static Defects}
\label{sec3b}

The electronic contribution to the splitting between the
energy levels of TLS is determined by the difference between
the two eigenvalues of the self-energy matrix~\cite{MW}
\begin{equation}
\label{selfenergy}
\left(\begin{array}{*{2}{c}}
V_3\delta_{++} & V_1\delta_{+-} \\ V_1\delta_{-+} &
-V_3\delta_{--}
\end{array}\right).
\end{equation}
The components of this matrix are expressed in terms of the
scattering matrix ${\cal T}$ corresponding to the potential
$U\left({\bf r}\right)$ introduced in Eq.\ (\ref{ranpot}) as
\begin{mathletters}
\begin{eqnarray}
\label{eq20}
\delta_{++}= - & 2i& \sum_i\int\frac{d\epsilon}{2\pi}
\int d{\bf r} \int d{\bf r}^{\prime} G_{\epsilon}
\left({\bf r}+ {\bf b}\right) \nonumber \\
& \times & {\cal T}_{\epsilon}
\left({\bf r}-{\bf r}_i,{\bf r}^{\prime}-{\bf r}_i\right)
G_{\epsilon}\left({\bf r}^{\prime}+ {\bf b}\right),
\end{eqnarray}
and similarly for $\delta_{--}$. The off-diagonal term
$\delta_{+-}$ is
\begin{eqnarray}
\label{eq20a}
\delta_{+-}=- & 2i & \sum_i\int\frac{d\epsilon}{2\pi}
\int d{\bf r} \int d{\bf r}^{\prime} 
G_{\epsilon}\left({\bf r}+ {\bf b}\right) \nonumber \\ 
& \times & {\cal T}_{\epsilon}\left({\bf r}-{\bf r}_i,
{\bf r}^{\prime}-{\bf
r}_i\right) G_{\epsilon}\left({\bf r}^{\prime}- {\bf
b}\right).
\end{eqnarray}
\end{mathletters}
The factor of $2$ in front is due to summation over channels
-- the two orientations of the real spin of the conduction
electrons. The resulting electronic contribution to the
splitting is
\begin{equation}
\label{eq19}
\Delta^2=V_3^2\left[{\mathrm{Re}}
\left(\delta_{++}-\delta_{--}\right)\right]^2+4V_1^2
\left({\mathrm{Re}}\,\delta_{+-}\right)^2.
\end{equation}

After integrating over $\epsilon$ and approximating
$\left|{\bf r}_i\pm {\bf b}\right|= r_i\pm {\bf b}\cdot {\bf
n}_i$, ${\bf n}_i={\bf r}_i/r_i$, the distribution function
for $\Delta$ is given by an expression analogous to Eq.\
(\ref{eq6}):
\begin{equation}
\label{eq20b}
{\cal P}\left(\Delta\right)=2\Delta\int\prod_i\frac{d{\bf
r}_i}{\cal V} \delta\left(\Delta^2-\sum_{ij}\frac{
K_1f_if_j+\widetilde{K}_3h_ih_j}{2} \right),
\end{equation}
where $h_i=h\left({\bf r}_i\right)= \displaystyle{\frac{\sin
2p_Fr_i}{\left(p_Fr_i\right)^3}} \sin \left(2p_F {\bf b}
\cdot {\bf n}_i\right)$, and constants $K_1$, $K_3$, and $t$
were defined above in Eqs.\
(\ref{Ks},\ref{t}). $\widetilde{K_3}$ is related to $K_3$
via $K_3=\left(2p_Fb\right)^2\widetilde{K_3}$.

Following the technique used in the preceding section, the
distribution function can be represented as an integral over
a Lagrange multiplier $\mu$ and two ``Hubbard-Stratonovich''
variables $\lambda_1$ and $\lambda_3$,
\begin{eqnarray}
\label{eq21}
{\cal P}\left(\Delta\right)=2 & \Delta& \int
\frac{d\mu}{2\pi}e^{i\mu\Delta^2}
\int\frac{d\lambda_1 d\lambda_3}{2\pi i\mu
\sqrt{K_1\widetilde{K}_3}}
\exp\left\{i\frac{\lambda_1^2}{2\mu K_1} \right. \nonumber \\
& + & \left. i\frac{\lambda_3^2}{2\mu \widetilde{K}_3} -\frac{4\pi
c}{3p_F^3}u\left(\lambda_1,\lambda_3\right)\right\},
\end{eqnarray}
where the function $u$ is defined by an expression analogous
to Eq.\ (\ref{eq10}),
\begin{equation}
\label{eq21a}
\frac{4\pi c}{3p_F^3}u\left(\lambda_1,\lambda_3\right) =
c\int d{\bf r}\left[1-\exp\left\{-i\lambda_1f\left({\bf
r}\right) -i\lambda_3h\left({\bf r}\right)\right\}\right].
\end{equation}
Integrating over orientations of $\mathbf{n}$, and
decoupling fast oscillations in $f$ and $h$, we arrive, in
the approximation $\sin 2p_Fb\approx 2p_Fb$, at an analogue
of Eq.\ (\ref{eq11}):
\begin{eqnarray}
\label{eq21b}
u\left(\lambda_1,\lambda_3\right)=
\int_0^\infty\frac{d\zeta}{\zeta^2}
\int_0^{2\pi}\frac{d\varphi}{2\pi}
& & \left[1 - \frac{\sin\left(\lambda_3
\widetilde{b}\sin\varphi
\zeta\right)}{\lambda_3\widetilde{b}\sin\varphi \zeta}
\right. \nonumber \\
& \times &  \exp\left\{-i\lambda_1\cos\varphi 
\zeta\right\}\Big],
\end{eqnarray}
where $\widetilde{b}=2p_Fb$ and
$\zeta=1/\left(p_Fr\right)^3$. Integration over $\varphi$
now gives
\begin{equation}
\label{eq21c}
u\left(\lambda_1,\lambda_3\right)=
\frac{2\widetilde{b}\left|\lambda_3\right|}{\pi}
\!\int_0^{\infty}\!\!\frac{d\zeta}{\zeta^2}
\left[1-\frac{\left|\lambda_1\right|\sin\zeta}
{\left|\lambda_3\right|\widetilde{b}} {\mathrm{K}}_{-1}
\left(\frac{\zeta\left|\lambda_1\right|}{\widetilde{b}\left|
\lambda_3\right|}\right) \right],
\end{equation}
where $\mathrm{K}_{\nu}$ is the modified Bessel
function. Completing the remaining integration over $\zeta$,
and rescaling $\widetilde{b}\lambda_3\rightarrow\lambda_3$
so that $\widetilde{K}_3$ in Eq.\ (\ref{eq21}) is replaced
with $K_3=\widetilde{b}^2\widetilde{K}_3$, we obtain
\begin{equation}
\label{eq22}
u\left(x,y\right)=\frac{x^2} {\left|y\right|} \ln\frac{
\left|y\right|+\sqrt{x^2+y^2}}{\left|x
\right|}+\sqrt{x^2+y^2}.
\end{equation}
Using polar coordinates $\left(\lambda,\theta\right)$ in the
$\left(\lambda_1,\lambda_3\right)$ plane, introducing
another set of polar coordinates $\left(\rho,\psi\right)$ in
the $\left(\lambda, \mu\right)$ plane and integrating over
$\rho$ we obtain the following integral representation for
the distribution function:
\begin{equation}
\label{eq22a}
{\cal P}\left(\Delta\right)=
\!\int_0^{2\pi}\!\!\!\frac{d\theta}{\pi^2\Delta}
\frac{K\left(\theta\right)}{\sqrt{K_1K_3}}{\mathrm{Re}}\,
\frac{\partial}{\partial \gamma}
\left. \!\int_0^{\pi/2}\!\!\!\!\!\frac{d\psi}{i\gamma\sin
2\psi-1}\right|_{\gamma=\frac{\gamma\left(\theta\right)}{\Delta}}.
\end{equation}
The final result obtained after integrating over $\psi$,
together with the definition of $\gamma\left(\theta\right)$,
has been presented in Eq.\ (\ref{eq23}).

\subsection{Higher Order Contributions.}

The scaling ansatz used in Eq.\ (\ref{eq30a}) depends
crucially on the shape of the distribution function~${\cal
P}$ in the region of small splitting~$\Delta$. Thus, the
applicability of the above analysis hinges on whether the
perturbative calculation of~$\Delta$ in the preceding
subsection is sufficient for~$\Delta$'s smaller than
$T_K$. Only the lowest order diagram in Kondo coupling has
been retained in the calculation so far, and we now turn to
the consideration of higher order contributions.

These higher order contributions can be conveniently
separated into two classes. The diagrams belonging to the
first class (Fig.~\ref{fig1}c) correspond to the
renormalization of the Kondo scattering amplitude with
parquet diagrams, which collect the leading order
logarithmic terms. They produce corrections to~$\Delta$
which are smaller by powers of the Kondo coupling, and,
importantly, unlike the Kondo scattering amplitude itself,
do not contain any logarithmically divergent terms. These
contributions can therefore be neglected.

Indeed, the contributions of the diagrams belonging to the
first class is typified by that of the diagram in
Fig.~\ref{fig1}b:
\begin{eqnarray}
\label{nextorder}
\delta\Delta_{\alpha} & = & \left(\nu g\right)\left(\nu
J\right)^2 \sum_i \frac{\pi E_F}{\left(p_Fr_i\right)^3} 
\cos\left(2p_Fr_i\right) S^{\alpha}_i 
\nonumber \\
& \times &
\left[\ln\frac{E_F}{v_F/r_i}+\mathrm{Const.}\right]
\end{eqnarray}
This expression is valid as long as
\begin{equation}
\label{ineq}
v_F/r_i < T_K, 
\end{equation}
which, for $T_K/E_F\sim 10^{-4}$ would be violated only at
unphysically small concentrations $c/p_F^3 < 10^{-12}$. It
is seen that this contribution does not contain any
uncontrolled logarithmic divergences. The reason is that the
sum of all diagrams of this type can be written in the form
of Eq.\ (\ref{eq4}), where the bare scattering amplitude $J$
is replaced with the renormalized amplitude
$J_R\left(\epsilon,\epsilon\right)$. The renormalized
amplitude is singular at small energies, but the singularity
is integrable, and after the integration over energies in
Eq.\ (\ref{eq4}) it only manifests itself in finite
logarithmic terms like the first term in square brackets in
Eq.\ (\ref{nextorder}). The inequality (\ref{ineq}) ensures
that the contribution of these terms is small in the
parameter $\left(\nu J\right)\ln\frac{E_F}{v_F/d} \ll 1$,
and can be ignored.

In contrast, the diagrams of the second class correspond to
a further renormalization of the scattering amplitude by
particle-hole pairs, and retain the usual logarithmically
divergent factors. These terms were previously analyzed
perturbatively using the renormalization group (RG) approach
in Ref.~\onlinecite{Zaw}, where it was established that in
anisotropic models the effect of these terms may be to
renormalize downwards the effective splitting {\em at the
scale just above $T_K$}. We argue that this analysis cannot
be straightforwardly extended through the crossover region
into the low-temperature regime. Instead, we show that the
effective splitting at temperatures below the Kondo
temperature can be deduced from the lowest order
perturbative result on the basis of universality properties
of the two models under consideration, the one- and
two-channel Kondo models. We employ the language of the
underlying Anderson model, as it affords a unified
description of the high- and low-energy
regimes~\cite{Hewson}.

The simplest diagram of the second class is shown in
Fig.~\ref{fig1}d. These graphs correspond to linear 
(in~$\Delta$) screening of the splitting by the Kondo
interaction. In the perturbative renormalization group
analysis by Vlad\'{a}r and Zawadowski~\cite{Zaw} it was
found that the splitting tends to be renormalized downwards,
and in the case of strong anisotropy, at least one component
of the splitting may be renormalized significantly,
\[
\Delta_{x\mathrm{, eff}}\left(T\ge T_K\right) \sim
\Delta_x\left(\frac{\nu J_1}{4\nu J_3} \right)^{1/4\nu J_3}.
\]
However, the perturbative analysis in Ref.~\onlinecite{Zaw}
cannot be extended to energy scales below $T_K$. An
extrapolation of the perturbative RG results from the energy
scale $E\ge T_K$ to predict the values of~$\Delta$ at $T\ll
T_K$ (Ref.~\onlinecite{vonDelft&Co}) is unjustified because
splitting is a relevant operator in the RG sense.

Let us consider first the isotropic one-channel magnetic
case. The perturbative renormalization group calculation of
Ref.~\onlinecite{Zaw}, although formulated in terms of the
orbital two-channel Kondo model, can be transferred, with
minor modifications, to the one-channel case as well because
of the identical structures of the corresponding
high-temperature perturbation expansions. When graphs of the
type shown in Fig.~\ref{fig1}b are neglected, the splitting
has the literal meaning of an external magnetic field
${\mathbf h}$ acting on the impurity.  The unrenormalized
impurity Green function has the form
\begin{equation}
{\cal G}=\left(\omega-\epsilon_d + {\mathbf
h}\cdot\bbox{\sigma}\right)^{-1},
\end{equation}
where $\epsilon_d$ is the energy of the singly occupied
impurity state. The ground state of the model can be
described as an effective Fermi liquid, in which the Green
function of the impurity spin retains its form under strong
renormalization. The fully renormalized ${\cal G}$ will
contain additional self-energy contributions~\cite{Hewson},
\begin{equation}
{\cal G}=\left[\omega-\epsilon_d + i\Gamma/z +
\bbox{h}\cdot\bbox{\sigma} -
\hat{\Sigma}\left(\omega,\bbox{h}\right)\right]^{-1} ,
\end{equation} 
where $\Gamma$ is the width of the Kondo resonance. The
self-energy
$\hat{\Sigma}=\Sigma_0+\bbox{\Sigma}\cdot\bbox{\sigma}$ is
expanded at small $\omega$ and $\mathbf{h}$ as
\begin{eqnarray}
\Sigma_0\left(\omega,\bbox{h}\right) & = & \Sigma_{00} +
\left(1-\frac{1}{z}\right)\omega + iO\left(\omega^2\right),
\nonumber \\ \bbox{\Sigma} & = & \Sigma^{\prime}_h \bbox{h},
\end{eqnarray}
where $\Sigma^{\prime}_h$ and $z$ are the renormalization
constants, and $\Sigma_{00}$ is the constant term which, in
the case of symmetric Anderson model, is equal to
$\epsilon_d$ ensuring that the Kondo resonance is centered
at the Fermi energy.

Factoring out the quasiparticle weight $z$, the {\em
effective} field ${\bf h}_{\mathrm{eff}}$ takes the form
\begin{equation}
\label{wilson}
\bbox{h}_{\mathrm{eff}}=z\left(1-\Sigma^{\prime}_h\right)\bbox{h}.
\end{equation}
The perturbative RG calculation of renormalized splitting in
Ref.~\onlinecite{Zaw} is equivalent to computing the same
prefactor $\displaystyle{R\left(\omega\right)=
z\left(1-\Sigma^{\prime}_h\right)\equiv
\frac{1-\partial\Sigma/\partial
\bbox{h}}{1-\partial\Sigma/\partial\omega}}$ at a {\em
finite frequency} $\omega$ (see Eqs.\ (3.5)-(3.10) in the
second of Refs.~\onlinecite{Zaw}). The frequency dependence
of both derivatives in this region is logarithmic, so that,
with logarithmic accuracy, frequency can be identified with
energy scale in the RG sense. Such scale-dependent
quantities determine the properties of the system at
temperatures of the order of the energy scale.

The perturbative RG analysis cannot be continued beyond some
intermediate scale $E_0 \gtrsim T_K$, and its usefulness for
determining low-temperature properties relies on the
assumption that further renormalization from $E\sim E_0$ to
$E\ll T_K$ does not change the value of the renormalized
quantity in any essential way. This assumption may be
violated when renormalization of relevant operators is
considered, as is indeed the case in the models considered
here.

At zero temperature, the prefactor in Eq.\ (\ref{wilson}) is
just the universal Wilson ratio $R$
(Ref.~\onlinecite{Hewson}).  Substituting its known value,
$R=2$, we obtain a seemingly counterintuitive result that,
despite downwards renormalization of splitting at $E>T_K$,
the splitting is actually increased at $E\ll T_K$ by a
factor of 2 compared to its bare value. Of course, at zero
temperature the role of the weak effective magnetic field
acting on the impurity is to polarize the Kondo screened
complex (as long as $\Delta \ll T_K$), and
$\bbox{h}_{\mathrm{eff}}$ induces a ``splitting'' only in
this sense.

To explain the non-monotonic behavior of the effective
splitting as a function of energy scale (or temperature),
one should note that, in the Anderson model, the
renormalization of the splitting is proportional to the
ratio of the impurity magnetic susceptibility $\chi$ to
$C/T$, where $C$ is the impurity specific heat. This ratio
is a non-monotonic function of temperature near $T_K$,
essentially because of the maximum in the temperature
dependence of $C$ at $T\sim T_K$.

The Anderson model is less well suited for discussion of
anisotropic Kondo models. Nevertheless, anisotropy can be
modeled, at the price of additional potential scattering in
the corresponding s-d Hamiltonian, by introducing
spin-dependent renormalization constants $z_{\alpha}$ and
$\Sigma^{\prime}_{\alpha\beta}$. Coupling anisotropy is an
irrelevant operator (in both one-channel and two-channel
spin-$1/2$ Kondo models, see, {\em e.g.}
Ref.~\onlinecite{PangCox}), and the invariance of the Wilson
ratio demands that
\begin{equation}
2=z_{\uparrow}\left(1-\Sigma^{\prime}_{\uparrow\uparrow}
\right)=
z_{\downarrow}\left(1-\Sigma^{\prime}_{\downarrow\downarrow}
\right)= \sqrt{z_{\uparrow}z_{\downarrow}}
\left(1-\Sigma^{\prime}_{\uparrow\downarrow}\right).
\end{equation}

The choice of only two different renormalization constants
$z_{\uparrow}$ and $z_{\downarrow}$ corresponds to
$J_1=J_2\ne J_3$ in the Kondo effective Hamiltonian. If the
field now is chosen in the $xy$ plane
($h_{\uparrow\downarrow}$ in the above notations), the
corresponding value of the self-energy at zero temperature
is still
\[h_{\uparrow\downarrow\mathrm{eff}}=h_{\uparrow\downarrow}
\sqrt{z_{\uparrow}z_{\downarrow}}
\left(1-\Sigma^{\prime}_{\uparrow\downarrow}\right)=
2h_{\uparrow\downarrow}.\] The temperature dependence of
this self-energy may be quite non-monotonic, as the
transverse susceptibility deviates strongly from the
free-spin value at temperatures above the specific heat
maximum.

Let us turn now to the two-channel Kondo case. The above
analysis cannot be transferred verbatim because the ground
state is not a Fermi liquid, and $T^*$ does not have the
meaning of the renormalized Fermi-liquid quasiparticle
self-energy~\cite{tsvelik}. In particular, the
renormalization factor $R\left(\omega\right)$ can no longer
be identified with the Wilson ratio of an Anderson-like model.
Nevertheless, the salient features of this analysis
survive. Once again, we can identify the self-energy
contribution of Fig.~\ref{fig1}a with an external field
${\bf h}$ acting on the impurity. The effective field in the
high-temperature regime is renormalized downwards, in the
anisotropic case strongly~\cite{Zaw}. However, in the
low-temperature regime the weight of the quasiparticle
excitations for which this effective field represents the
self-energy is zero, so that this self-energy term does not
define any physical low-temperature energy scale. It has
been demonstrated recently~\cite{tsvelik,schofield} that
{\em impurity} thermodynamics in the low-temperature regime
can be described in terms of three vector and one scalar
Majorana quasiparticles. An external field ${\bf h}$
generates a self-energy contribution for the scalar Majorana
fermion $\Sigma_M=h^2/2\pi T_K=T^*$ which is universal apart
from its dependence on
$T_K$~\cite{PangCox,SacraSchlott,sengupta}. The
universality~\cite{remark2} ensures that $T^*$ is controlled
by the unrenormalized value of $h$, namely the bare
splitting~$\Delta$.

\section{Discussion.}
\label{disc}
 
The results of our model calculation of the splitting
between the states of magnetic impurities and non-magnetic
two-level systems (TLS) in the presence of spin disorder can
be summarized as follows. First, the main feature of the
distribution functions of the splittings is their strong
asymmetry. Formally the distribution functions in Eqs.\
(\ref{eq17}) and (\ref{eq24}) do not even possess finite
first moments. This is a manifestation of the fact that all
the moments of the splittings are dominated by disorder
configurations with one or more scatterers located very
close to the dynamic impurity.  In real systems the shortest
possible separation between the dynamic impurity and the
nearest scatterer is determined by the lattice
spacing. Therefore one has to introduce a short distance,
lattice cut-off into the coordinate integration in Eq.\
(\ref{eq10}). In the presence of such a cutoff all the
moments are dominated by distances of the order of the
cut-off, and thus, for example, the average rms splitting
$\displaystyle{\sqrt{\left\langle\Delta^2\right\rangle}}$ is
of the order of $E_F\left(Jg\nu^2\right) \sqrt{c/p_F^3}$ for
the magnetic case or $E_F\left({\rm
max}\left\{V_1,V_3p_Fb\right\}\right)t\nu \sqrt{c/p_F^3}$
for the non-magnetic TLS case. These values are larger than
the typical ones by a factor of $\sqrt{p_F^3/c}\sim
\left(p_Fd\right)^{3/2}\gg 1$. This confirms the result
obtained independently by D.~Cox~\cite{cox1}.

Second, since the splitting is the difference between two
eigenvalues of a random Hermitian $2\times 2$ matrix, at
small splittings one observes the equivalent of level
repulsion leading to a vanishing probability density to
observe zero splitting. In the non-magnetic TLS case the
matrix is real and symmetric, and the suppression is linear
rather than quadratic.

The asymmetry of the distribution functions has important
implications for the study of the effects of spin disorder
on the behavior of Kondo impurities in the crossover and
low-temperature regimes. Because of the build-up of
many-particle correlations at low temperatures, the
effective splitting between the levels of magnetic
impurities or TLS, $\Delta_{\rm eff}\left(T\right)$, can be
reduced at intermediate temperatures $T\sim T_K$ compared to
the bare value~$\Delta$ (Refs.~\onlinecite{Zaw,MF}). Since
the distribution of splittings is very asymmetric, with the
root-mean-square splitting much larger than the typical
value, a proper renormalization analysis would necessarily
treat the full distribution rather than just the first few
moments. To what extent the distributions found here
preserve their shape under renormalization is an open
question.

Nevertheless, the zero-temperature behavior of the
splitting, or, more precisely, of the corresponding
self-energy terms, is dictated by the universality
properties of the one- and two-channel Kondo models, and can
be extracted directly. This, in turn, made it possible to
derive a corresponding temperature dependence of conductance
for a collection of two-level systems. We find
(cf. Ref.~\onlinecite{WAM}) a $T^{\alpha}$ behavior with
$\alpha\approx .84$ in contrast to the $T^{1/2}$ dependence
observed experimentally~\cite{RB,RBLvD,vonDelft&Co}. The
concentration of the TLS which must be degenerate (in the
absence of disorder) in order to sustain the two-channel
Kondo interpretation is also found to be unphysically
large. Both these arguments suggest that the two-channel
Kondo model does not provide a consistent interpretation of
the zero-bias anomalies observed in Ref.~\onlinecite{RB}.

In conclusion, we have computed the distribution functions
of splittings of magnetic impurities and non-magnetic
two-level systems (TLS) induced by disorder scattering of
conduction electrons. In the magnetic case these splittings
only appear if the disorder breaks time-reversal symmetry,
{\it i.e.} if the disorder is itself magnetic. In the
non-magnetic case the degeneracy between the levels of a TLS
is due to geometric symmetry about its center, and therefore
is strongly broken by any type of disorder. We find that the
probability distribution of splittings vanishes as a power
law at small splittings, making nearly degenerate impurities
a rarity. The typical values of splittings are found to be
smaller than the average estimated previously in
Ref.~\onlinecite{WAM}. However, even in quite clean systems
such as the ones studied in the experiments of
Ref.~\onlinecite{RB}, the broad profile of the distribution
of splittings results in a temperature dependence of the
conductance which is substantially different from the
square-root law expected in the absence of
disorder. Consequently, experimental observation of the
square-root temperature and voltage dependences may not be a
reliable indicator of two-channel Kondo physics.

\section*{Acknowledgments}
The authors would like to thank B.~Altshuler, N.~Cooper,
J.~von~Delft, D.~S.~Fisher, M.~P.~A.~Fisher, H.~Li, Y.~Meir,
A.~Schofield, B.~Simons, J.~Ye and G.~Zarand for valuable
discussions. I.~S. acknowledges partial financial support
from NSF grant \# DMR 94-16910 at Harvard University, and
also the hospitality of the NEC Research Institute where
part of this work was performed.

\appendix

\section{}
\label{suscept}

\begin{figure}
\begin{center}
\leavevmode
\epsfxsize=3in
\epsfbox{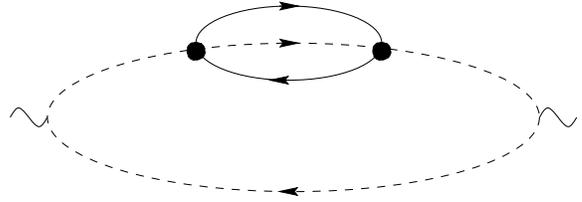}
\bigskip
\caption{\label{fig6} The lowest order logarithmic
contribution to susceptibility in the magnetic Kondo
problem. The wavy lines correspond to the external magnetic
field.}
\end{center}
\end{figure}

To illustrate the small effects of spin disorder on the
standard logarithmic terms in the perturbation-theory
expansion, we consider the well-known lowest-order
logarithmic contribution to the spin susceptibility given by
the diagram shown in Fig.~\ref{fig6}.

As in other similar logarithmic diagrams, there appears a
set of terms due to the off-diagonal in spin index parts of
the Green function.  The corresponding analytical expression
is

\begin{eqnarray}
\label{eq2.40}
& \displaystyle{\frac{\delta\chi\left(T\right)}{\chi_0}}&  =  
\frac{J^2}{3}
\int \! d\epsilon_1 d\epsilon_2 
\!\left[3\rho\left(\epsilon_1\right)\rho
\left(\epsilon_2\right)\! -
\sum_{j=1}^3\sigma_j\left(\epsilon_1\right)\sigma_j
\left(\epsilon_2\right)\right]
\nonumber \\
& \times & \!\!\!\left[
\frac{1-\tanh\frac{\epsilon_1}{2T}\tanh
\frac{\epsilon_2}{2T}}
{\left(\epsilon_1-\epsilon_2\right)^2}-2T
\frac{\tanh\frac{\epsilon_1}{2T}-\tanh
\frac{\epsilon_2}{2T}}
{\left(\epsilon_1-\epsilon_2\right)^3} 
\right], \nonumber \\ \!\!\!\!\!
\end{eqnarray}
where $\sigma_j\left(\epsilon\right)$ is the electron spin
density of states at the impurity site, and $\chi_0$ is the
susceptibility of a free spin. Since the average
$\langle\sigma_j\rangle =0$, there is no positive definite
contribution to the logarithmic integral similar to that
coming from the first term $3\rho\left(\epsilon_1\right)
\rho\left(\epsilon_2\right)$. Moreover, absence of diffusive
behavior for $\sigma_j$ (due to the absence of an equivalent
of the particle-number conservation law which enforces the
universal diffusion pole in the density correlations) leads
to the absence of ``fine structure" in the correlator
$\langle\sigma_j\sigma_j\rangle$ at scales smaller than
$\tau_s^{-1}$. In the model of isolated impurities the
effect of the second term in Eq.\ (\ref{eq2.40}) reduces to
a small correction to the coefficient of the leading
logarithm~\cite{comment2}.

Indeed, in the lowest order in concentration of magnetic
defects ${\bf S}_a$, the local spin density of states is
given by
\begin{equation}
\label{eq.a1}
\bbox{\sigma}\left(\epsilon\right)= -\frac{g}{\pi}\sum_a
\left(\frac{\pi\nu}{p_Fr_a}\right)^2 {\rm Im}
e^{-2i\left(p_F+\epsilon/v_F\right)r_a}{\bf S}_a,
\end{equation}
where ${\bf r}_a$ are the coordinates of the defects. Using
$\langle S_a^iS_b^j\rangle=
\displaystyle{\frac{1}{3}}\delta_{ab} \delta_{ij}$, we find
\begin{eqnarray}
\label{eq.a2}
\left\langle\sigma_i\left(\epsilon_1\right)\sigma_i
\left(\epsilon_2\right)\right\rangle & = & 
\frac{cg^2}{\pi^2}\int d{\bf r}
\left(\frac{\pi\nu}{p_Fr} \right)^4
\sin\left[2\left(p_F+
\epsilon_1/v_F\right)r\right] \nonumber \\
& \times & \sin\left[2
\left(p_F+\epsilon_2/v_F\right)r\right] \nonumber \\  = 
\pi^2\nu^2\left(\pi g\nu\right)^2 &
\displaystyle{\frac{c}{p_F^3}} &
\left[4+\frac{\epsilon_1+\epsilon_2}{E_F}-
\frac{\left|\epsilon_1-\epsilon_2\right|}{E_F}\right].
\end{eqnarray}

Substituting the last expression into Eq.\ (\ref{eq2.40}) we
see that the second and the third term in square brackets
produce irrelevant constants and terms small as $T/E_F$,
while the first term results simply in a correction to the
effective density of states $\delta\rho^2/\nu^2 = -
\displaystyle{\frac{4\pi
c}{3p_F^3}}\pi^4\left(g\nu\right)^2$.

\section{}
\label{param}

The Kondo temperature in the TLS model is given
by~\cite{Zaw}
\begin{equation}
\label{kondotemp}
T_K\approx E_F\sqrt{\left(\nu J_1\right)\left(\nu
J_3\right)} \left(\frac{J_1}{4J_3}\right)^{\frac{1}{4\nu
J_3}},
\end{equation}
where $J_1$ and $J_3$ are electron-TLS coupling constants,
and $\nu$ is the local conduction electron density of
states. Assuming a two-channel Kondo interpretation of the
anomalies, the Kondo temperature is estimated in
Ref.~\onlinecite{RB} to be between $5\mathrm{K}$ and
$10\mathrm{K}$. The choice of the bare coupling values $\nu
J_3\approx .2$ and $\nu J_1\approx .007$, which is adopted
in the calculations leading to the graphs in
Figs.~\ref{pofdelta} and \ref{condplot}, corresponds to the
Kondo temperature of $T_K\approx 8.2K$. The respective
values of the couplings $V_3$ and $V_1$ in Eq.\ (\ref{eq18})
are $V_3\approx .74$, and $V_1\approx .0012$, corresponding
to the dimensionless distance between the TLS minima set at
$2p_Fb\approx .15$.

The nature of defects in quenched vacuum-deposited films is
not well understood, and is likely to vary depending on the
details of a particular experimental setup. As a rough
measure of disorder, the transport mean free path near the
opening of the constriction is estimated in
Ref.~\onlinecite{RB2} to be $\ell_{\mathrm{tr}}\sim
30\mathrm{nm}$ in {\em unannealed} samples. The mean free
path is related to the scattering phase shifts $\eta_l$ and
concentration of defects $c$ via~\cite{Mahan}
\begin{equation}
\label{pl}
p_F\ell_{\mathrm{tr}} =\frac{p_F^3}{4\pi c}
\frac{1}{\sum_{l=1}l\sin^2\left(\eta_l-\eta_{l-1}\right)},
\end{equation}
where $l$ denotes angular momentum channels.  The transport
mean free path in {\em annealed} samples is shown in
Ref.~\onlinecite{RB2} to be close to $300{\mathrm{nm}}$,
{\em i.e.} it increases by a factor of 10 upon
annealing. Therefore, most of the disorder in the
constriction is likely to be caused not by substitution
impurities which cannot anneal, but by localized structural
inhomogeneities in stressed deposited films. Spatially
extended defects, such as dislocations and grain boundaries,
are unlikely to be a major cause of scattering for the same
reason.

The coefficients $K_1$ and $K_3$ in Eq.\ (\ref{eq23}) cannot
be expressed in terms of the transport mean free path
alone. Their computation requires the knowledge of the phase
shifts $\eta_l$ [which would make it possible to infer the
concentration from Eq.\ (\ref{pl})]. Even allowing for an
independent measurement of defect concentration (one not
available for the Ralph-Buhrman samples), information about
phase shifts is still needed if more than one of them is
non-zero.

It is seen from Eq.\ (\ref{eq23}) that the scale for the
typical values of~$\Delta$ is set by the combinations
$\left(c/p_F^3\right)\sqrt{K_i}$, $i=1,3$, and, for small
$\eta_l$, it is linear in both $c$ and $\eta_l$. Together
with Eq.\ (\ref{pl}), this implies that, for a fixed value
of $\ell_{\mathrm{tr}}$, smaller values of $\eta_l$ result
in a broader distribution function ${\cal
P}\left(\Delta\right)$. For example, the choice of
$\left|\sin\eta\right|=1/6$ leads to typical values of $T^*$
exceeding $T_K$. Conversely, if scatterers are strong
($\eta_l\sim 1$), fewer of them are needed to produce the
same value of $\ell_{\mathrm{tr}}$, and the distribution
function is narrowed, implying smaller typical
disorder-induced splittings.

It should be noted that because of the Friedel sum rule,
small values of {\em all} $\sin\eta_l$ are possible only in
the case of neutral defects. If defects are charged, at
least one of the phase shifts cannot be small.

A typical neutral defect is a combination of a
self-interstitial and a neighboring vacancy (a so-called
Frenkel defect). We are not aware of any studies of whether
such defects are more or less common in vacuum-deposited
films than simple non-equilibrium vacancies, which are
charged, and thus scatter strongly. However, the formation
energy of such neutral defects in Cu is $2$ to $3$ times
larger than the formation energy of a
vacancy~\cite{physmet}. Consequently, we concentrate on the
non-equilibrium vacancies as the dominant source of
scattering. Since one of the aims of this work is to explore
whether the two-channel Kondo interpretation of the
Ralph-Buhrman zero-bias anomalies can be compatible with the
presence of the static scattering in their samples, the
assumption that the bulk of disorder is caused by strongly
scattering non-equilibrium vacancies is also quite
appropriate. It leads to the smallest values of splittings,
and is, therefore, the most favorable for the two-channel
Kondo interpretation of the experiments in
Ref.~\onlinecite{RB}.

Scattering by vacancies in Cu has been studied by several
techniques (experimental, numerical, and their combination)
in Ref.~\onlinecite{lodder}. The results of the fitting
procedure combining experimental and numerical techniques
quoted in Ref.~\onlinecite{lodder} give for the differences
between the vacancy and the host phase shifts the following
set of values: $\eta_0\approx .92$, $\eta_1\approx -.7$, and
$\eta_{l\ge 2}=0$.

\section{}
\label{randomtk}

Fluctuations of the Kondo temperature $T_K$ due to static
disorder can be addressed in the same framework used for the
calculation of the distribution of splittings. Indeed, in
the leading logarithmic order, the Kondo temperature can be
inferred from the scaling relation~\cite{Zaw}
\begin{equation}
\label{scaling}
\frac{d\psi}{\Phi\left(\psi\right)}=-d\ln D,
\end{equation}
where $D$ is the renormalized band cut-off, $\psi$ is a
homogeneous degree one function of dimensionless couplings
$J_i\nu$, and $\Phi$ is a homogeneous function of degree two
which depends on $\psi$ and bare values of the
couplings. The effect of disorder is contained in the
energy-dependent density of states $\nu\left(D\right)$. The
homogeneity of both $\psi$ and $\Phi$ allows one to make a
substitution $\psi\left(J\nu\right)
\rightarrow\psi\left(J\overline{\nu}\right)$, where
$\overline{\nu}$ is the average density of states, while
transferring the energy dependence of $\nu$ to the r.h.s. of
Eq.\ (\ref{scaling}). The implicit equation for $T_K$ takes
the form
\begin{equation}
\label{tk}
\int_0^1\frac{d\psi}{\Phi\left(\psi\right)}=
\int_{T_K}^{D_0} \frac{\left[\nu\left(D\right)+
\nu\left(-D\right)\right]dD}{2\overline{\nu}D},
\end{equation}
where $D_0\sim E_F$ is the unrenormalized bandwidth, and the
boundary condition $\psi\left[J\left(D_0\right)\right]=0$
was used. Following the consideration in
Section~\ref{sec3b}, the correction to the density of states
as a function of energy $\epsilon$ is expressed as
\begin{equation}
\label{dos}
\delta\nu\left(\epsilon\right)=
\nu\left(\epsilon\right)-\overline{\nu}=
-\frac{t\overline{\nu}}{2}\sum_{j}\frac{{\mathrm{Im}}\,
e^{2i\left(p_F+\epsilon/v_F\right)r_j}}
{\left(p_Fr_j\right)^2},
\end{equation}
where $\bbox{r}_j$ are the coordinates of random scatterers.
Substituting the above expression into Eq.\ (\ref{tk}) we
obtain the following condition for the disorder-induced
shift $\delta T_K$:
\begin{equation}
\ln\frac{T_K+\delta T_K}{T_K}+\sum_j\frac{
t\sin \left(2p_Fr_j\right)}{\left(p_Fr_j\right)^2} 
\,{\mathrm{ci}} \left(\frac{2T_Kr_j}{v_F}\right)=0.
\end{equation}
The argument of the integral cosine ${\mathrm{ci}}$ is small
as $\left(T_K/E_F\right)\left(p_F^3/c\right)^{1/3}\sim
10^{-3}$, where the values $T_K/E_F\approx 10^{-4}$ and
$c/p_F^3\approx 10^{-4}$ are used (see
Section~\ref{experiment}). It can therefore be approximated
as ${\mathrm{ci}}\left(2T_Kr_j/v_F\right)\approx
-{\mathrm{\bf C}}-\ln\left(2T_Kr_j/v_F\right)$, where
${\mathrm{\bf C}}$ is the Euler constant. Typical values of
$\delta T_K/T_K$ are thus seen to be of the order of
$t\left(c/p_F\right)^{2/3}\ln\left[
\left(T_K/E_F\right)\left(c/p_F^3\right)^{-1/3}\right]$, and
we estimate $\delta T_K/T_K\sim 10^{-2}$. Therefore random
variations of $T_K$ {\em due to static disorder} can be
safely neglected for the parameter regime of
Ref.~\onlinecite{RB}.

Intrinsic variation of $T_K$ among different randomly formed
TLS is, of course, a possibility. However, it has been
established in the main text that all TLS have to be
degenerate in order for the two-channel Kondo interpretation
to have a chance of succeeding. Thus one must assume that
the TLS are formed by the same non-random mechanism, again
excluding large variations of $T_K$.


\begin{references}
\bibitem{elastic}Effects of elastic strains, and intrinsic
asymmetry of TLS are beyond the scope of the present work. A
broad distribution of intrinsic splittings would render the
two-channel Kondo effect unobservable, and for the purposes
of analyzing the Ralph-Buhrman experiments we assume that
electronic disorder is the only source of asymmetry. While
it is not implausible that, in the absence of disorder, the
TLS are naturally degenerate~\cite{vonDelft&Co}, generation
of splittings by elastic strains, which are almost certain
to be present in vacuum deposited films before annealing,
deserves a separate study. In neglecting their contribution
altogether we can obtain only the lower bounds on the
characteristic values of splittings.
\bibitem{channels}In the case of TLS, channel quantum
numbers in Eq.\ (\ref{nck}) would refer to physical spin,
while spin quantum numbers correspond to parity about the
center of the TLS, or, more precisely to linear combinations
of angular momentum eigenstates which are either even or odd
under parity~\cite{Zaw}. A two-dimensional subspace of all
such states has been shown in Ref.~\onlinecite{Zaw} to be
the only subspace relevant in the renormalization group
sense.
\bibitem{note4}Since the representation in Eq.\ (\ref{eq18})
is more convenient for the computational purposes, we do not
present the corresponding $H_s$ contribution explicitly.
\bibitem{comment2}It is easy to show that the distribution
of the spin-disorder contribution to $\delta\chi$ possesses
a strong asymmetry similar to that in the distribution of
the splittings, so that the average
$\langle\delta\chi\rangle$ is dominated by the tails of its
distribution. However, since a physical measurement of
susceptibility directly corresponds to measuring the average
$\langle\delta\chi\rangle$ we do not analyze its
distribution in detail.
\end{references}
\end{document}